\documentclass[]{interact}

\usepackage{epstopdf}

\usepackage{caption} 
\usepackage{subcaption}

\usepackage{epigraph}
\usepackage[backend=biber, style=authoryear-comp, natbib=true]{biblatex}
\DeclareLanguageMapping{english}{english-apa}
\usepackage[hidelinks]{hyperref}
\usepackage[frozencache=true,cachedir=.]{minted}

\usepackage[table,xcdraw]{xcolor}
\usepackage{pdflscape}

\theoremstyle{plain}

\theoremstyle{definition}

\theoremstyle{remark}

\usepackage{verbatim}
\usepackage[utf8]{inputenc}

\usepackage{adjustbox}
\usepackage{graphicx}
\usepackage{listings}
\usepackage{comment} 
\usepackage{enumitem} 
\usepackage{multirow} 
\usepackage{colortbl} 
\usepackage{hhline} 

\usepackage{xcolor}
\definecolor{class_null}{HTML}{737373}
\definecolor{class_a}{HTML}{ffffe5}
\definecolor{class_b}{HTML}{fff7bc}
\definecolor{class_c}{HTML}{fee391}
\definecolor{class_d}{HTML}{fec44f}

\colorlet{punct}{red!60!black}
\definecolor{background}{HTML}{EEEEEE}
\definecolor{delim}{RGB}{20,105,176}
\colorlet{numb}{magenta!60!black}

\lstdefinelanguage{json}{
    basicstyle=\normalfont\ttfamily,
    numbers=left,
    numberstyle=\scriptsize,
    stepnumber=1,
    numbersep=8pt,
    showstringspaces=false,
    breaklines=true,
    frame=lines,
    backgroundcolor=\color{background},
    literate=
     *{0}{{{\color{numb}0}}}{1}
      {1}{{{\color{numb}1}}}{1}
      {2}{{{\color{numb}2}}}{1}
      {3}{{{\color{numb}3}}}{1}
      {4}{{{\color{numb}4}}}{1}
      {5}{{{\color{numb}5}}}{1}
      {6}{{{\color{numb}6}}}{1}
      {7}{{{\color{numb}7}}}{1}
      {8}{{{\color{numb}8}}}{1}
      {9}{{{\color{numb}9}}}{1}
      {:}{{{\color{punct}{:}}}}{1}
      {,}{{{\color{punct}{,}}}}{1}
      {\{}{{{\color{delim}{\{}}}}{1}
      {\}}{{{\color{delim}{\}}}}}{1}
      {[}{{{\color{delim}{[}}}}{1}
      {]}{{{\color{delim}{]}}}}{1},
}

\begin{document}

\articletype{RESEARCH ARTICLE}

\title{Choriented Maps: Visualizing SDG Data on Mobile Devices}

\author{
\name{Viktor Gorte and Auriol Degbelo\thanks{CORRESPONDING AUTHOR: Auriol Degbelo. Email: auriol.degbelo@uni-muenster.de}}
\affil{Institute for Geoinformatics, University of Münster}
}

\vspace{2cm}

\maketitle

\begin{abstract}
Choropleth maps and graduated symbol maps are often used to visualize quantitative geographic data. However, as the number of classes grows, distinguishing between adjacent classes increasingly becomes challenging. To mitigate this issue, this work introduces two new visualization types: choriented maps (maps that use colour and orientation as variables to encode geographic information) and choriented mobile (an optimization of choriented maps for mobile devices). The maps were evaluated in a graphical perception study featuring the comparison of SDG (Sustainable Development Goal) data for several European countries. Choriented maps and choriented mobile visualizations resulted in comparable, sometimes better effectiveness and confidence scores than choropleth and graduated symbol maps. Choriented maps and choriented mobile visualizations also performed well regarding efficiency overall and performed worse only than graduated symbol maps. These results suggest that the use of colour and orientation as visual variables in combination can improve the selectivity of map symbols and user performance during the exploration of geographic data in some scenarios. 

\end{abstract}




\section{Introduction}
\label{sec:introduction}
Sustainable Development Goals (SDG) have been suggested by the United Nations to meet the greatest challenges faced by the planet and the importance of developing techniques to communicate their outcomes has been acknowledged in previous work (e.g. \cite{Kent2020,Smith2020,Yuan2020}). This article is concerned with the visualization of SDG information on mobile devices. Mobile devices could indeed become the medium of choice to inform the public opinion about progress on SDG since an increasing number of people use them for information-gathering tasks (e.g. 57 percent of adults in the U.S., see \cite{Walker2019}). The research question addressed in this article is: how to best visualize quantitative geographic data (e.g. SDG data) on mobile devices for comparison purposes (i.e. \textcolor{black}{data comparison tasks})?

\textbf{Why comparison:} There is currently little empirical evidence of user needs in the context of SDGs indicators communication (e.g. the taxonomy of open data user needs from \cite{degbelo2020needs} appears here too coarse). In the absence of such evidence, the focus on comparison in this work was guided by two factors: theoretical and practical relevance. From the theoretical point of view, comparison is a basic task that is relevant for both exploratory (i.e. derive hypotheses) and confirmatory analysis (i.e. test assumed hypotheses). Comparison is indeed present in several taxonomies of visualization tasks, for instance, \citep{Brehmer2013, Schulz2013a,roth2013empirically}. From the pragmatic perspective, comparison has practical relevance in the context of SDG data because it can help data consumers put values in perspective (e.g. how a country fare in the European context or the more global context of all countries worldwide). Visualizations optimized for, or supporting comparison, if they are available on mobile devices can be used in addition to reports such as \citep{Nations2020}, to inform the public about progress on SDG. 

\textbf{Why choriented:} Choropleth and graduated symbol maps are well known and the most frequent in the context of open data geovisualization (see e.g. \cite{degbelo2020datascale}). Nonetheless, as the number of classes grows, distinguishing between adjacent classes becomes increasingly challenging. This work proposes maps that use both colour and orientation as variables to encode geographic information to mitigate this issue. \textcolor{black}{Redundant symbolization with orientation (i.e. the use of orientation in combination with other visual variables) has been suggested previously in the literature.} For instance, \textcite{Bertin1983} indicated that variations in orientation are least selective for the representations of areas (e.g. lakes, countries), and thus should be used in combination with a selective variable (Page 93). Likewise, \textcite{MacDonald1999} recommended using colour in conjunction with other visual variables (e.g. shape, orientation, texture) for an effective presentation on computer displays (Page 25). While the combination of orientation with other variables has been suggested, the extent to which it is effective still needs to be empirically tested and documented. This work is an attempt to address this gap. The name `choriented' is derived from a combination of `choropleth' + `orientation', as choriented maps supplement choropleth maps with orientation. Choriented mobile visualizations are an optimization of choriented maps for mobile devices (see Figure \ref{fig:app_visualizations}). Contrary to existing work on the effectiveness of the combination of visual variables on multivariate maps (e.g. \cite{Nelson1999,Nelson2000,elmer2013symbol}), choriented maps use the combination to improve selectivity on univariate maps (i.e. the encoding of one attribute). The contributions of this article are twofold:
\begin{itemize}
    \item An open-source prototype to visualize SDG indicators on mobile devices. The key features include a flexible, dynamic generation of user interface components based on the SDG data provided as input, good visualization rendering performance, and the translation of the codebase into native mobile applications for multiple platforms.
    \item An evaluation of four strategies to visualize SDG indicators with respect to four types of comparison tasks. The newly proposed choriented maps and choriented mobile visualizations resulted in comparable, sometimes better effectiveness and confidence scores. Choriented maps and choriented mobile visualizations performed well also regarding efficiency overall and performed worse only than graduated symbol maps.

\end{itemize}

\section{Related work}
\label{sec:relatedwork}
\color{black}
This section briefly introduces previous work in three areas relevant to the current article: visual variables and redundant symbolization, the visualization of SDG data and the design of visualizations on mobile devices. Existing work on mapping SDG has synthesized cartographers' knowledge, provided tools for map generation, and investigated user preferences, but more empirical research is needed to pinpoint the merits of different visualization types, particularly on mobile devices. 


\subsection{Visual variables and redundant symbolization}
\color{black}
Visual variables are properties of visual symbols that can be used to encode information. \textcite{Bertin1983} originally proposed seven visual variables and these were extended to a list of 12 in \citep{maceachren2004maps,MacEachren2012visual}: location, size, colour hue, colour value, colour saturation, orientation, grain, arrangement, shape, fuzziness (a.k.a. crispness) and resolution.  The combination of variables to depict a single attribute has been termed `redundant symbolization' \citep{Roth2017a,White2017}, `redundant stimulus dimensions' \citep{dobson1983visual}, `redundant encoding' \citep{Chun2017}, `redundant cues' \citep{Nelson2020}, or `cartographic redundancy' \citep{Cybulski2018} in the literature. A related notion to redundant symbolization, but slightly different from it is that of bivariate/multivariate mapping, i.e. the encoding of two or more data attributes concurrently into a single symbolization mechanism, in order to communicate relationships between the attributes (see \cite{Nelson2020}). It should be noted that redundant symbolization and bi/multivariate mapping are not mutually exclusive. For instance, since bi/multivariate maps are information-dense products, redundant symbolization may be used to reinforce the visual interpretation of one of their thematic variables as illustrated in \citep{Nelson2020}.

\textcite{Shortridge1982} was an early work studying the impact of redundancy in symbolization on the discrimination of map labels. They reported that redundant visual cues on map labels (i.e. the combination of lettering size, case, boldness and dot size) yield faster response times and a greater percentage of correct responses. \textcite{dobson1983visual} investigated the effect of using size+colour(value) for map reading tasks on graduated symbol maps and reported performance improvements of the redundancy condition over the use of size alone. More recently, \textcite{Cybulski2018} compared maps using colour+size to maps using colour only to reduce change blindness on animated maps. The redundant encoding yielded improved effectiveness in their study. Colour in their work was a combination of hue and value (they used a diverging scheme from green to red). Finally, \textcite{Chun2017} investigated the impact of the conjunctions colour(value)+position and colour(value)+size and reported no significant effects of the redundant encoding on either accuracy or judgment speed. The size variable was encoded as length, angle and area in his study. He then concluded that ``more information in the form of redundant variations of [colour] value can potentially enhance design without weakening the integrity of graphic perception". 

As stated in Section \ref{sec:introduction}, empirical tests are currently needed to learn about the effectiveness of the conjunction colour+orientation and this is the focus of this article. The three visual variables of interest in the experiment presented later are size (variations in the length, area, or volume of a symbol); colour value (light or dark variations of a single hue); and orientation (the direction or angle of rotation of a symbol).







\color{black}

\subsection{Visualization of SDG data}
SDG data is used by an increasing number of tools. Mobile applications that deal with SDG data aim to inform the broad population about the concept of SDG. This can be seen in Samsung's latest release of the Samsung Global Goals App\footnote{\url{https://play.google.com/store/apps/details?id=com.samsung.sree}, last accessed: May 13, 2021.} that intends to inform the broad population and provides means to donate to United Nations programs. Table \ref{tab:tools_table} shows existing visualization tools and some of their features as of May 2021\footnote{Chart (a.k.a. diagram or graph) is used in the table in line with \citep{Kraak2020} to denote a representation of non-geographic attribute and temporal data patterns.}. There are two key takeaways from the table. First, most visualizations support comparison explicitly, and this is confirmatory evidence of the importance of this task in the context of SDG. Second, tools that offer maps as a visualization focus mostly on choropleth maps. While choropleth maps are valuable, there is more than one way of visualizing a given dataset. Empirical studies informing about the merits of other types of visualizations may inform developers of SDG visualizations about \textit{when} they should turn to alternatives (in addition to guidelines from the literature). The study presented later in this article is a contribution along these lines.

Scientific contributions related to the visualization of SDG data to date have provided a synthesis of the available knowledge \citep{Kraak2018,Kraak2020}, a tool for map generation \citep{Gong2019}, and results from the investigation of user preferences \citep{gosling2019, Pirani2019}. Since developers of web software and web-based visualization are not always cartography savvy, making cartographers' knowledge about map making and use accessible to a broader audience is needed. To that end, \textcite{Kraak2018} have discussed the cartographic workflow (i.e. a series of steps including data transformation, selection of visual variables, choice of map types) in relation to the sensemaking of SDG indicator data. The International Cartographic Association (ICA) and the United Nations have partnered to publish a comprehensive set of guidelines for mapping geographic datasets related to the SDGs \citep{Kraak2020}. The book introduces principles of map design and use, best practices, and explains how different mapping techniques support the understanding of the SDGs. As regards tools, \textcite{Gong2019} proposed `SDG Viz', a web-based tool to help create different types of maps to explore SDG data. The types of maps supported by SDG Viz include: single/multi-colour choropleth maps, proportional symbol maps, pattern area maps, 3D maps, and animated maps. As to user preferences, \textcite{gosling2019} were concerned with finding out the most appropriate projection to represent small island developing states. SDG 1 (Poverty) was considered as a theme, and the feedback from users suggested that the Interrupted Goode Homolosine projection is promising for the representation task. \textcite{Pirani2019} compared a choropleth map, a contiguous cartogram, and a tilemap for the visualization of SDG 5 (Gender equality) data. Users' comments have indicated familiarity with and acceptance of the choropleth map, but the tilemap showed the highest performance on informational tasks. They also called for further studies investigating the impact of different types of thematic maps on users' perception, a call that is answered partly later in this article (Section \ref{sec:userstudy}).

\begin{table}[H]
\centering
\caption{Existing online tools to visualize SDG data. }
\label{tab:tools_table}
\resizebox{\textwidth}{!}{%
\begin{tabular}{|l|l|l|l|l|l|} 
\hline
Tools                                                                   & \begin{tabular}[c]{@{}l@{}}Visualization \\Type \end{tabular} & \begin{tabular}[c]{@{}l@{}}Explicit Spatial \\Encoding \end{tabular} & \begin{tabular}[c]{@{}l@{}}Supports \\Comparison \end{tabular} & \begin{tabular}[c]{@{}l@{}}Responsive\\Design \end{tabular} & Notes                                                                                        \\ 
\hline
SDGs \& me                                                                & Chart         & NO                                                                   & YES                                                           & YES                                                         & No maps                                                                                      \\ 
\hline
\begin{tabular}[c]{@{}l@{}}World Development \\Indicators \end{tabular} & Chart         & NO                                                                   & YES                                                           & NO                                                          & No maps                                                                                      \\ 
\hline

SDGs Dashboard        & Chart \& Map           & YES                                                                  & YES                                                           & YES                                                         & \begin{tabular}[c]{@{}l@{}}Choropleth map only \end{tabular}        \\ 
\hline
\begin{tabular}[c]{@{}l@{}}SDG Index \\Dashboard \end{tabular}          & Chart \& Map           & YES                                                                  & YES                                                           & YES                                                         & \begin{tabular}[c]{@{}l@{}}Choropleth map only  \end{tabular}                   \\ 
\hline
\begin{tabular}[c]{@{}l@{}}Our World in Data \end{tabular} & Chart \& Map         & YES                                                                   & YES                                                           & YES                                                          & \begin{tabular}[c]{@{}l@{}}Choropleth map only. The tool is  \\not dedicated to SDGs, but has a SDG \\ tracker section for SDG data\end{tabular}                                                     \\ 
\hline
VizHub Healthdata                                                        & Chart \& Map    & YES                                                                  & YES                                                           & YES                                                          & \begin{tabular}[c]{@{}l@{}}Choropleth map only. The tool is  \\not focused on SDG, but has a section \\ for SDG health data visualization\end{tabular}  \\
\hline
\end{tabular}
}
\end{table}

\subsection{Visualization on mobile devices}
\textcite{Chittaro2006} documented the typical constraints facing visualizations on mobile devices: smaller screen size, lower resolution, different width/height ratio, and less powerful hardware. Additional constraints mentioned in \citep{Ricker2018,roth2019mobilefirst,Roth2018} include less reliable data connectivity, reduced bandwidth, variable environmental conditions as a result of physical mobility, limited battery life, and multi-touch post-WIMP interaction. Visualizations on mobile devices may be designed to support wayfinding and navigation (e.g. \cite{Burigat2013,Dillemuth2005,Meng2005,Cheung2009}) or the sensemaking of geographic data. The focus of this article is on the latter purpose. As \textcite{Lee2018} reminded, transferring recommendations for visualization development to mobile devices remains under-explored in the literature. A few recent works provide a starting point. \textcite{Du2018} proposed banded choropleth maps as an alternative to both small multiples and animation. Banded choropleth maps, \textcite{Du2018} argue, can use limited screen space more efficiently than small multiples, and preserve context more clearly than animation. Banded choropleth maps could thus help mitigate the issue of mobile devices' limited screen size (though it is slightly unclear if their empirical evaluation used mobile devices). \textcite{Brehmer2020} looked into the comparison of trends in datasets on mobile devices, and compared animation and small multiples for this task. They reported that users using small multiples completed the majority of the tasks in less time than those using animation. Also, those using small multiples were slightly less confident in their responses than those using animation. Contrary to the works above, the work uses interaction to assist the user during the sensemaking of SDG data.

\section{Prototype}
\begin{figure}[H]
  \begin{subfigure}[t]{.4\textwidth}
    \centering
    \includegraphics[width=\linewidth]{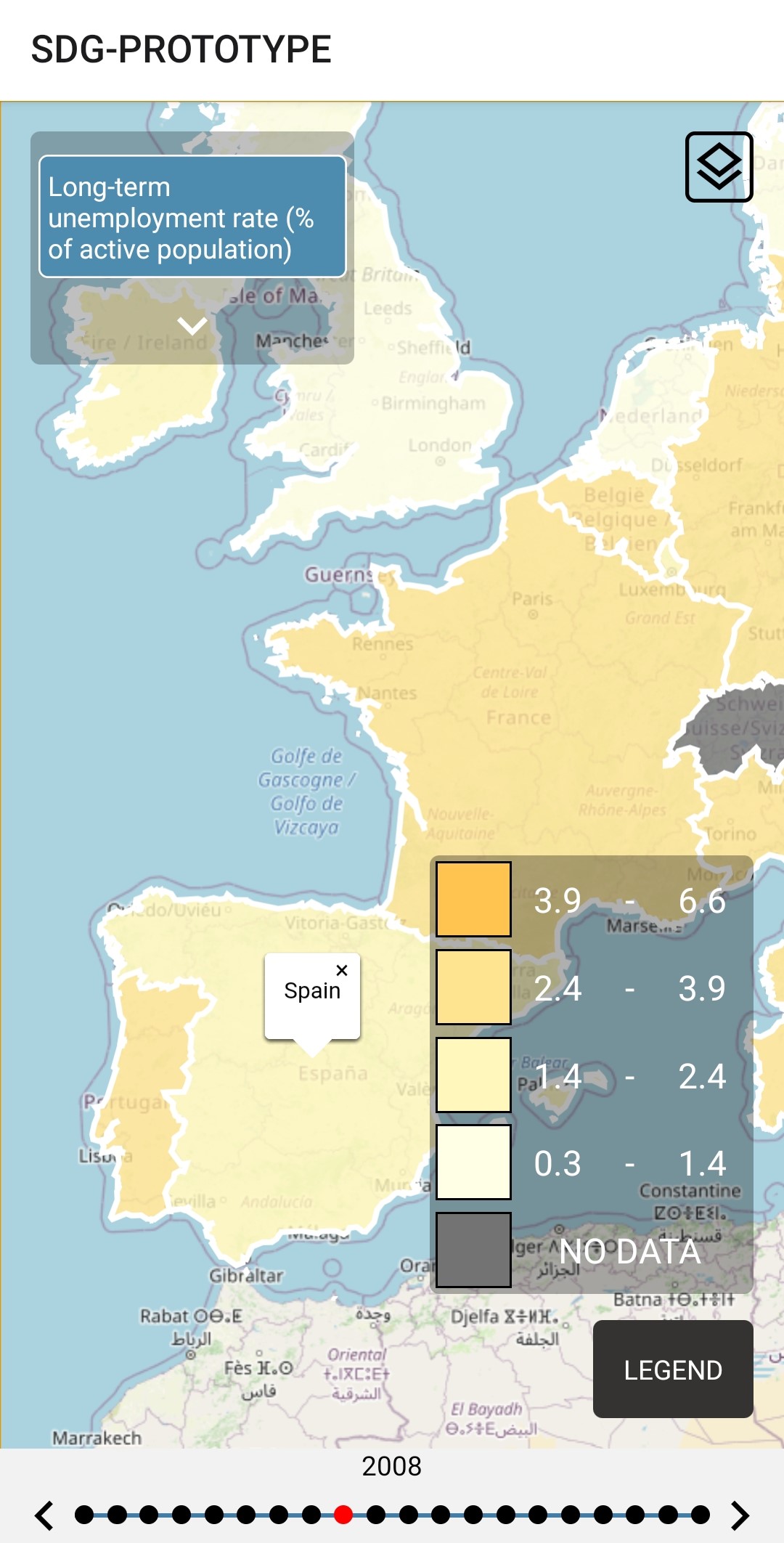}
    \caption{Choropleth Map}
    \label{fig:app_choropleth}
  \end{subfigure}
  \hfill
  \begin{subfigure}[t]{.4\textwidth}
    \centering
    \includegraphics[width=\linewidth]{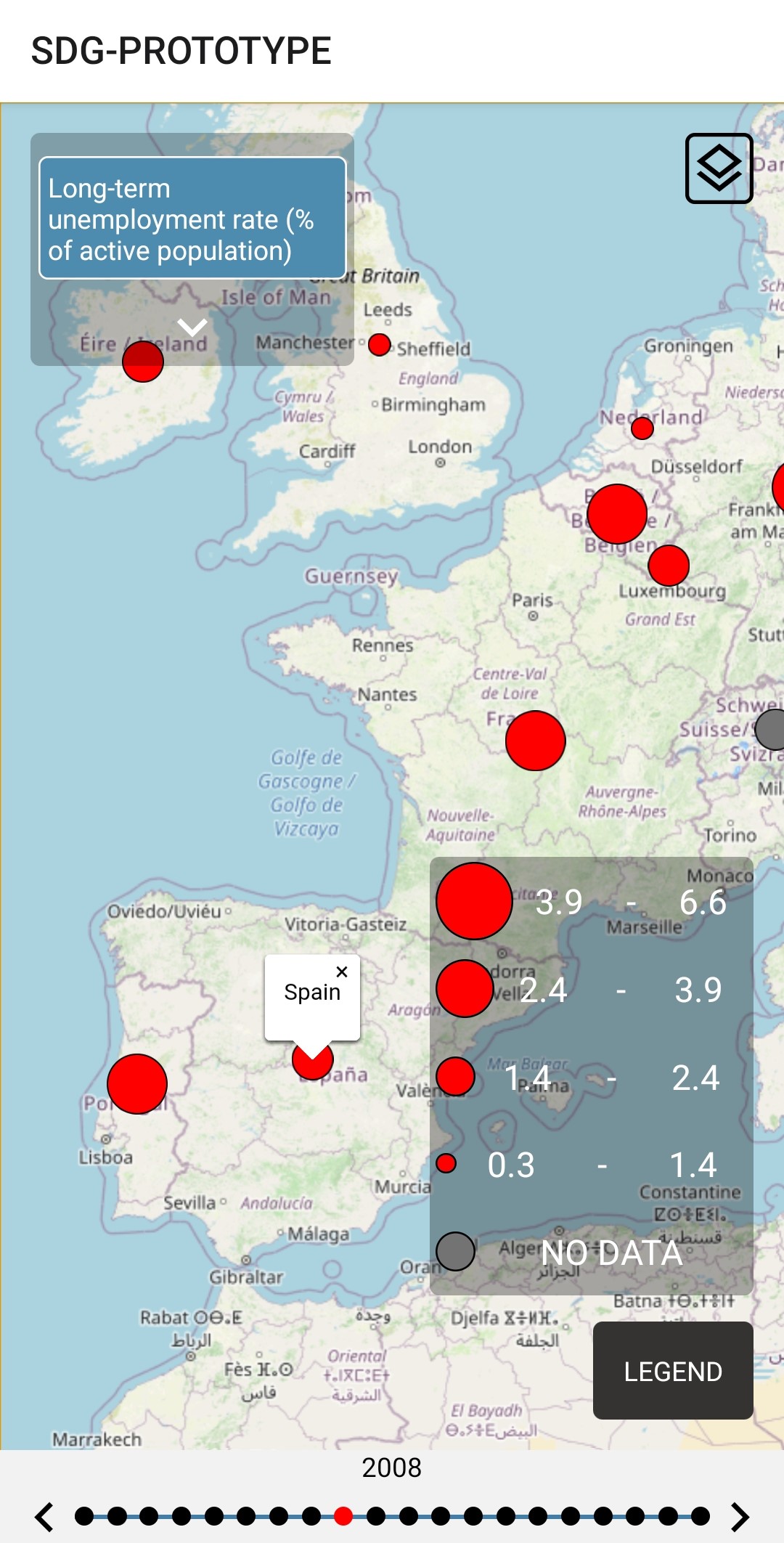}
    \caption{Graduated Symbol Map}
    \label{fig:app_gsm}
  \end{subfigure}

  \medskip

  \begin{subfigure}[t]{.4\textwidth}
    \centering
\includegraphics[width=\linewidth]{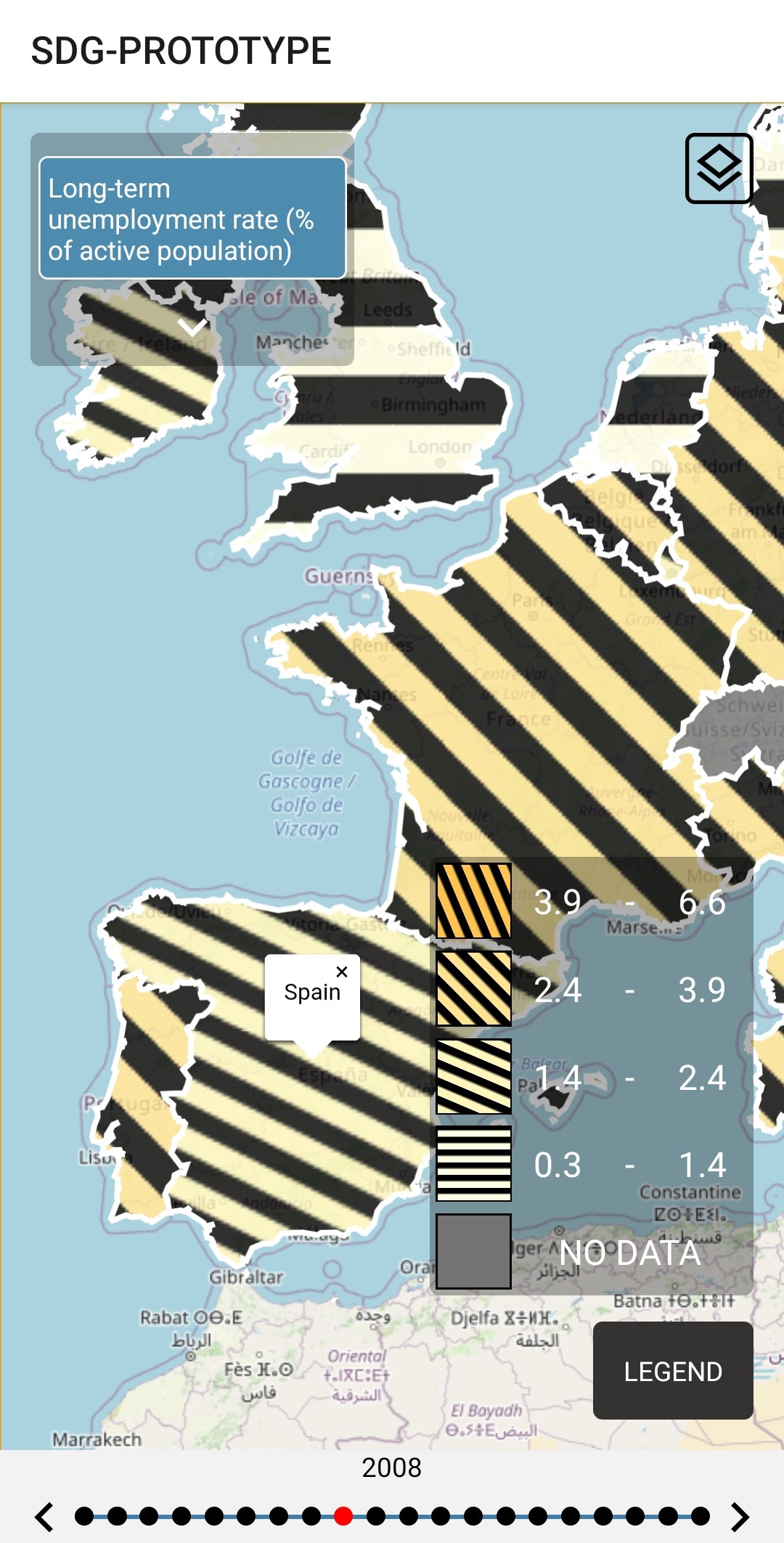}
    \caption{Choriented Map}
    \label{fig:app_chorientedmap}
  \end{subfigure}
  \hfill
  \begin{subfigure}[t]{.4\textwidth}
    \centering
    \includegraphics[width=\linewidth]{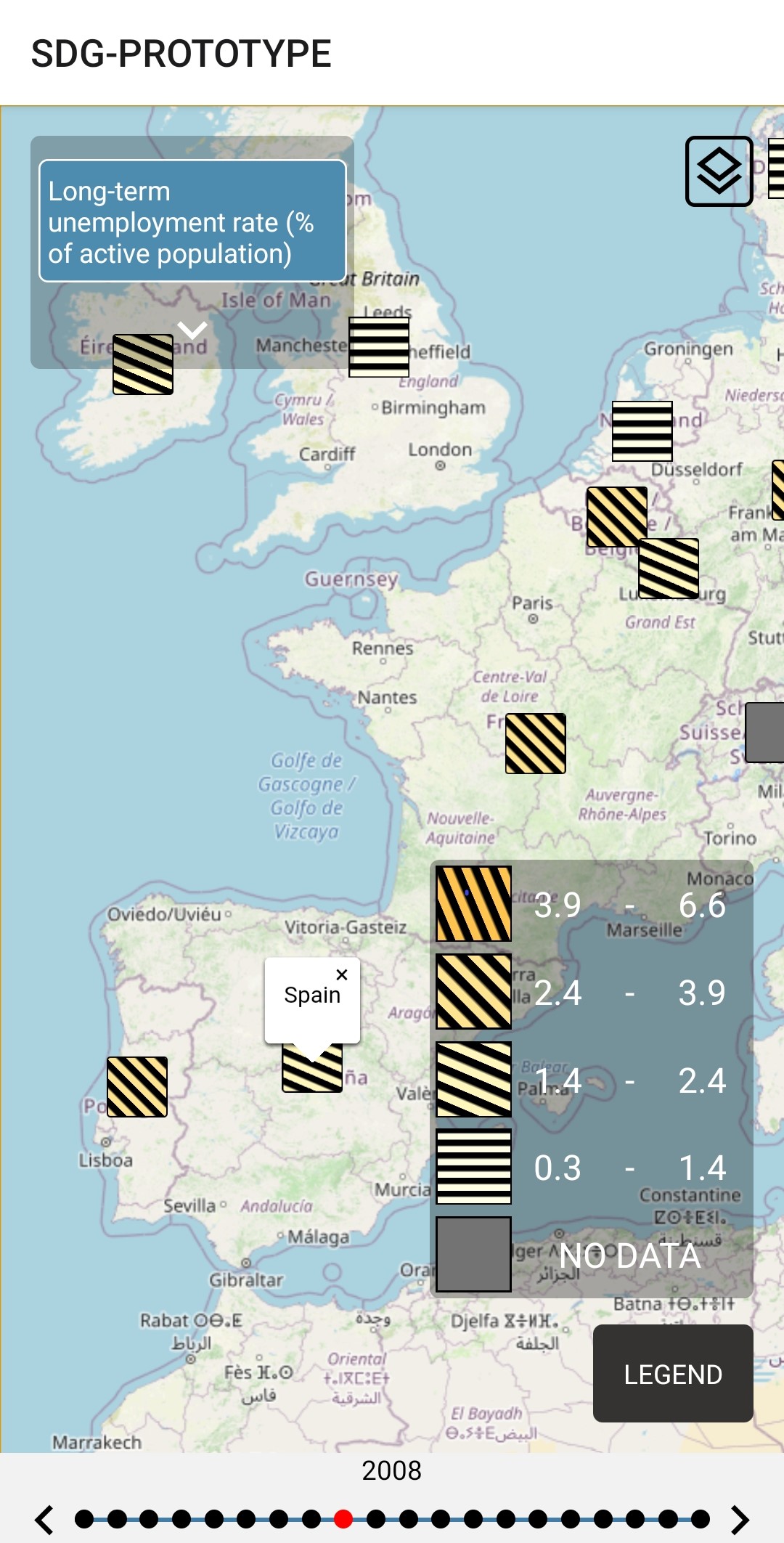}
    \caption{Choriented Mobile Map}
    \label{fig:app_chorientedmobile}
  \end{subfigure}
    \caption{Visualization types in the prototype application with extended legends.}
    \label{fig:app_visualizations}
\end{figure}

This work introduces a new type of map visualization called Choriented Map. It uses two visual variables to increase the perceptual selectivity of symbols. The used visual variables are \textit{colour value} and \textit{orientation}. This combination aims to improve comparison performance regarding attribute values of geographical entities in a map-based visualization on mobile devices. Similar to choropleth maps, Choriented Maps cover the shape of the geographic entity. The colour value is used to fill the entire shape of the geographic entity. \textcolor{black}{A sequential colour scheme \citep{brewer1994color,brewer1999color} is used, that is, colour values with darker colours represent higher values.} Orientation directions range from horizontal lines (lowest value) to vertical lines (highest value). An example of a Choriented Map with colour values and orientation lines can be seen in Figure \ref{fig:app_visualizations}c. Since Choriented Maps introduce a rather complex pattern that fills the geographic shape, potentially cluttering the whole device screen, a lessened version is introduced. The visualization is reduced to a square marker which is placed in the center of the geographic shape and contains the same geographic data representation as the standard Choriented Map. This reduced version is called Choriented Mobile Map and an example is shown in Figure \ref{fig:app_visualizations}d.

\subsection{Generating Choriented Maps}
Choriented Maps are based on colour values and orientation lines as visual variables. This combination produces a rather complex pattern that is not trivial to generate. This section presents two approaches of how to generate the orientation lines with a coloured background. The orientation of the striped pattern is dependent on the number of classes to be visualized and belongs to the interval $[0, 179]$ degrees. This range results from the fact that orientation lines between 180 and 360 degrees are indistinguishable from orientation lines between zero and 179 degrees. Orientations are assigned to classes with equal distances between zero and 179 degrees, where a lower value represents a lower class. Generating choriented maps can happen through the image approach, or the CSS (Cascading Style Sheet) approach. Both approaches are used in the prototype mobile application. The Choriented Map visualization utilizes the image approach and the Choriented Mobile Map visualization utilizes the CSS approach.

\begin{figure}[H]
    \centering
    \begin{subfigure}[b]{0.39\textwidth}
        \includegraphics[width=\textwidth, height=2in]{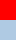}
        \caption{Minimal pattern image, original size is 16x40 pixels.}
        \label{fig:minimal_pattern_image}
    \end{subfigure}
    \begin{subfigure}[b]{0.39\textwidth}
        \includegraphics[width=\textwidth, height=2in]{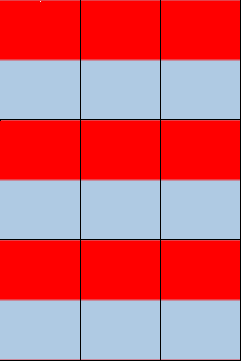}
        \caption{Striped pattern, black lines divide individual images.}
        \label{fig:enlarged_stripes_image}
    \end{subfigure}
        \caption{Generating a fill pattern from images}\label{fig:striped_pattern}
\end{figure}

The first approach for generating Choriented Maps utilizes images and is based on the support for fill-patterns for filling polygons with images. A map visualization tool that supports fill-patterns is Mapbox\footnote{\url{https://docs.mapbox.com/mapbox-gl-js/api/}, last accessed: May 14, 2021.}, which is also used in the prototype application. Fill-patterns are used for drawing polygons from repeating image patterns and are specified when adding a layer to the map. Therefore, filling a polygon with a striped pattern and a background colour requires an image that can be repeated on all sides to result in a pattern that fills the shape. Since memory and network access may be limited (mobile device limitations), a minimal pattern image is desirable. An example of a minimal pattern image for horizontal stripes can be seen in Figure \ref{fig:minimal_pattern_image}. Minimal pattern images can be created with online stripe generator tools\footnote{\url{http://www.stripegenerator.com/}, last accessed: May 14, 2021.} by specifying the background colour and the orientation of the lines. The minimal pattern image results in a larger image with a clear pattern by repeating the minimal pattern image in any dimension, which can be seen in Figure \ref{fig:enlarged_stripes_image}. The repetition process is automated by the Mapbox tool. 

The second approach for generating Choriented Map patterns relies on CSS instead of images. This is the reason why this approach is most suitable for generating custom HTML markers for web-based map visualization tools. This solution is used for visualizing the Choriented Mobile Map and utilizes the \textit{repeating-linear-gradient}\footnote{\url{https://developer.mozilla.org/en-US/docs/Web/CSS/repeating-linear-gradient()}, last accessed: May 14, 2021.} CSS function which creates images consisting of repeating linear gradients. Custom HTML markers are a widely supported feature in map visualization tools and replace map markers with HTML elements that can be styled with CSS. An example implementation of the styling of a custom HTML marker with repeating linear gradients, which results in a striped pattern with a specified background colour and black orientation lines with a 45-degree orientation with a width of 20 pixels, which can be seen in listing \ref{lst:stripes_css}.

\begin{lstlisting}[language=json,caption={CSS example for generating striped pattern.}, label={lst:stripes_css}]
.striped_pattern_45deg {
    width: 30px;
    height: 30px;
    background: repeating-linear-gradient(
        45deg,
        #fee391,
        #fee391 20px,
        black 20px,
        black 40px
    );
}
\end{lstlisting}

\subsection{Implementation}
The prototype application is designed to be easily extensible when it comes to features or datasets used. The application has been developed using the React-Native framework. React-Native was used because instead of rendering web content in a platform-dependent web view, the code base is converted into native components. This results in native mobile applications for multiple platforms from a common code base with better performance, more security, and better gesture support (e.g. tap, pinch, or spread) compared to hybrid mobile applications.

\subsubsection{Functionalities}
The application allows for the visualization of data by users providing GeoJSON datasets (countries' geometries) and a JSON dataset (given values for a SDG goal). The minimized UI component that users can interact with can be seen in the top left of Figure \ref{fig:app_choropleth}. To enable flexible integration of SDG data into the app, the UI component is not hardcoded but rather \textit{generated dynamically} based on the dataset to be visualized. The application also features layer selection button (or icon) to select the desired visualization type as shown in Figure \ref{fig:app_choropleth} (top right).  Whenever data is missing, for a specific geographic entity or time period, the geographic shape is rendered in grey (Figure \ref{fig:app_visualizations}). Using colour saturation to represent missing data is one of many techniques for representing absence in Cartography (see \cite{Robinson2019}), and was chosen for its simplicity to implement and interpret. Finally, the prototype application is a native mobile application and this enables smooth multi-touch screen interaction.

SDG data is released annually by the UN and therefore has a temporal aspect attached to it. The temporal component of the data is made accessible to the user by the introduction of temporal controls, which are always visible and can be seen at the very bottom of Figure \ref{fig:app_choropleth}. Each dot represents a selectable year for which an entry in the data file exists. The currently selected year is represented with a red dot, instead of a black dot for currently not selected years, and can be changed by using the arrows on the bottom left and right. The currently selected time frame is displayed on top of the controls. Changing the currently selected time frame results in a re-rendering of the visualizations on the map and a re-classification for all values. This UI component is dynamically generated as well, based on the available time frames for each goal in the data file. To ensure that participants do not confuse geographic entities with one another, clicking on a geographic entity renders a popup that contains the name of the geographic entity (Figures \ref{fig:app_choropleth} to \ref{fig:app_chorientedmobile}). \textcolor{black}{This popup is shown at the most distant internal point from the polygon outline (also called the `pole of inaccessibility'). The implementation was done using the JavaScript library Polylabel\footnote{\url{https://github.com/mapbox/polylabel}, last accessed: May 20, 2021.}.} The classification of the data was done according to the Jenks natural breaks classification method. Further details on the prototype (e.g. schema of the data to be uploaded) and installation instructions can be found on GitHub\footnote{\url{https://github.com/vgorte/SDG-Prototype}}.

\subsection{Performance}
\begin{figure}[H]
    \centering
    \includegraphics[width=\textwidth]{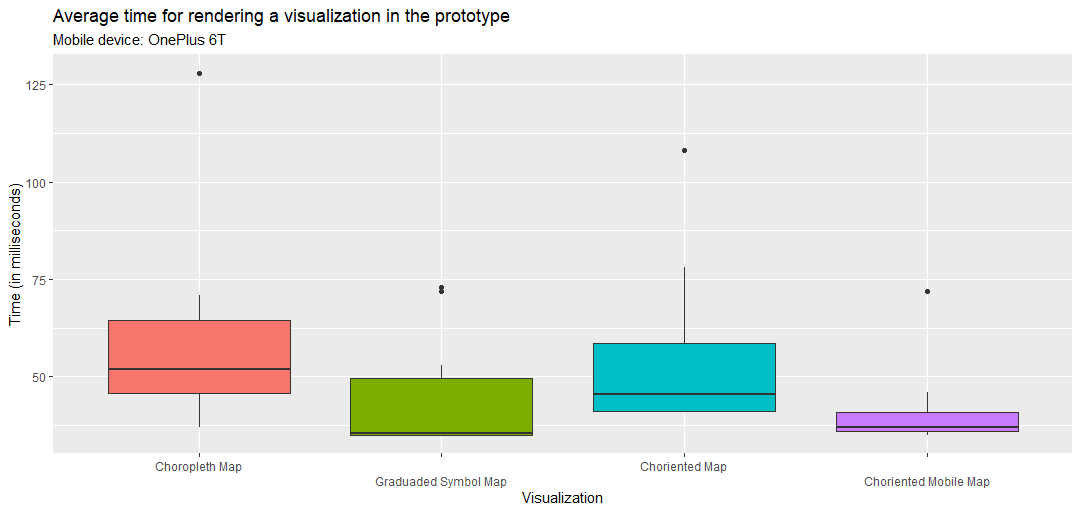}
    \caption{Benchmark of rendering ten visualizations per type in the prototype application.}
    \label{fig:benchmark_performance}
\end{figure}

To get an impression of the rendering performance, all four visualization types were compared. Figure \ref{fig:benchmark_performance} shows the results, averaged across ten rendering iterations. The figure shows that although mobile devices inherently come with less computing power compared to desktop devices, the prototype still performs well when it comes to rendering the visualizations. The measurements were done on a OnePlus 6t device. The rendering time is well below 100 milliseconds and this suggests that the prototype would be appropriate for operational usage. \textcolor{black}{In particular, an important takeaway from the figure is that choriented maps and choriented mobile maps are at least as good as choropleth maps, as far as visualization rendering time is concerned.}

\section{User study}
\label{sec:userstudy}
The goal of the study was to evaluate how well participants solve information-gathering tasks on mobile devices using the four map visualization types supported by the prototype.

\subsection{Variables} 
The independent variables of the study were: the map visualization types (Choropleth map, Graduated Symbol map, Choriented map, Choriented Mobile map) and the types of tasks (described below). The dependent variables included: efficiency (time needed to solve the tasks), effectiveness (number of correct answers), users' self-reported confidence about their answers \textcolor{black}{(measured on a five-point Likert scale for the statement `I am sure that my give answer is correct'), and their subjective impressions of the suitability of the four visualization types for solving the tasks}.

\subsection{Tasks}
As mentioned at the beginning of the article, the focus of this work was on comparison tasks. `Compare' is an objective primitive and denotes interactions that `determine the similarities and differences between two map features' \citep{roth2013empirically}. As discussed in \citep{gleicher2018consideratio} comparison is a multi-faceted concept. In particular, the target of the comparison (i.e. the set of items being compared) could be a source of ambiguity. In this work, the target of the comparison (i.e. map features being compared) is the individual countries. The comparison activity is \textit{mediated} by the map legends.   

The information-gathering tasks were defined as questions, based on previous work \citep{roth2013empirically,Sarikaya2018}. The tasks presented to the study participants focused on the attributes-in-space operand, which describes interactions with the mapped attributes to understand how characteristics of geographic phenomena vary in space \citep{roth2013empirically}, and the space-in-time operand, which describes interactions with the temporal component of the map to understand how geographic phenomena change over time \citep{roth2013empirically}. Each primitive operand was further divided into two groups, leading to four types of tasks for the participants: attributes-in-space clusters, attributes-in-space frequency, space-in-time trends, and space-in-time distribution. The attributes-in-space operand was divided into tasks that focus on the extraction of information about how often specific classes are represented in the visualization (frequency) \citep{Sarikaya2018}, and tasks that focus on the identification of geographic entities that belong to a common class (clusters) \citep{Sarikaya2018}. The space-in-time operand was divided into tasks, which aim at the extraction of information regarding high-level changes across the spatial and temporal dimensions of the data (trends) \citep{Sarikaya2018} and into tasks that focus on the extraction of information regarding the extent and frequency of value occurrences over time (distribution) \citep{Sarikaya2018}. The questions asked during the study are shown in Table \ref{tab:informationtasks}.

\begin{table}[]
\caption{Types of information-gathering tasks given to the participants during the study.}
\label{tab:informationtasks}
\resizebox{\textwidth+2cm}{!}{%
\begin{tabular}{|l|l|l|}
\hline
Operand       & \begin{tabular}[c]{@{}l@{}}High-level data \\ characteristics\end{tabular} & Questions                                                                                                                                                                                                                       \\ \hline
attributes-in-space   & Clusters                                                                   & \begin{tabular}[c]{@{}l@{}}- Which countries have the same life expectancy as Lithuania in 2004?\\ - Which countries have a higher area under organic farming than Italy in 2018?\end{tabular}                                  \\ \hline
attributes-in-space   & Frequency                                                                  & \begin{tabular}[c]{@{}l@{}}- How many countries have fewer people at risk of poverty or social exclusion than France in 2013?\\ - How many countries have the same long-term unemployment rate as Greece in 2012?\end{tabular}  \\ \hline
space-in-time & Trends                                                                     & \begin{tabular}[c]{@{}l@{}}- How did the number of people at risk of social exclusion in Germany change from 2005 - 2010?\\ - How did the area under organic farming in Portugal change from 2004 - 2009?\end{tabular}          \\ \hline
space-in-time & Distribution                                                               & \begin{tabular}[c]{@{}l@{}}- Which years show France and Germany with differing values for people at risk of poverty or social exclusion?\\ - Which years show Finland and Sweden having the same life expectancy?\end{tabular} \\ \hline
\end{tabular}
}
\end{table}

Another simple way of formally describing the tasks of this study is by using the taxonomy from \citep{Schulz2013a}. Following the notation suggested in \citep{Schulz2013a}, the current study is \{ exploratory, browse$|$similarities$|$discrepancies, clusters$|$frequency$|$trends$|$distribution, attrib(*time)$|$attrib(*structure), multiple \}. That is, the assumed high-level goal of all visualization tasks was an exploratory analysis, which is concerned with deriving hypotheses from an unknown dataset (the dataset was unknown to the participants before the study). The tasks involved carrying three actions: browsing the data\footnote{Browsing is defined here after \citep{Brehmer2013} as searching for the characteristics of known map elements.}, seeking similarities, and seeking discrepancies between map elements. The tasks aim to reveal one of four high-level SDG data characteristics: clusters, frequency, trends, or distribution as mentioned above. The target of the visualization task (i.e. on which part of the data the task is carried out) was the temporal features (how do attributes vary over time), and the structural features (how are attributes linked to each other) of the map elements (i.e. countries). Finally, multiple instances of the target (i.e. multiple time instants and relationships between countries) were considered during the study\footnote{\textcolor{black}{Comparison as defined in this work is hence different from the `compare' task examined in \citep{Somska-Przech2021}.}}.

As for the data, the focus was on European countries so that study participants could find their way around the map reasonably well and that we would have a dense neighborhood of countries with which to test how well the attribute information conveyed could be distinguished. The time frames were chosen to be as large as possible based on the available data, to increase the likelihood that changes of a detectable magnitude could be observed. The topics of the questions were varied throughout to avoid learning effects on a specific dataset. Four datasets were included: \textit{Dataset 1}: Area under organic farming (in \%; time frame: 2000 - 2018; SDG2: Zero hunger); \textit{Dataset 2}: People at risk of poverty or social exclusion (in \%; time frame: 2005 - 2018; SDG1: No poverty); \textit{Dataset 3}: Life expectancy (in years; time frame: 2000 - 2018; SDG3: Good health and well-being); \textit{Dataset 4}: Long-term unemployment rate (in \% of active population; time frame: 2000 - 2019, SDG8: Decent work and economic growth). \textcolor{black}{The data was retrieved from the World Bank Data Portal \& Tools.}

\subsection{Procedure}
The study was conducted remotely due to global social distancing restrictions at the moment of the experiment (\textcolor{black}{August 2020}). The experimenter (first author) and the participant communicated via webcams and microphones while sitting in adjacent rooms. The study was presented to participants in the form of an online questionnaire and consisted of two parts. First, a background questionnaire where general information about the participants and their mobile device usage was presented. In the second part of the study, the participants were given a mobile device that ran the prototype application and visualized SDG data (limited to European countries to reduce the information density on the screen). Then they went through the tasks. A task was a combination of a question and a visualization type. Preventing learning effects was achieved by using a balanced Latin square approach where a unique combination of the four task types and the four visualization types determined which visualization type a participant used to answer questions with the prototype. The visualization type and task type combinations were presented to the participants twice (repetition), that is, a participant answered the same type of question using two different datasets (see Table \ref{tab:informationtasks}). The online tool for presenting the study to the participants (i.e. LimeSurvey) measured the time the participants needed to answer a given question. This time was used for comparing the participants' performance. In total, the experiment had 16 participants x 4 visualizations x 4 tasks x 2 repetitions per task type = 512 trials. The study was pilot-tested and approved by the institutional ethics board. 

\subsection{Colour Scheme}
To provide a basis for comparison with future studies, the colour scheme used during the work is documented in this section (see also Table \ref{tab:colorused}). The colours were selected using a sequential theme from Colorbrewer \citep{Harrower2003}. The ``null'' class (grey) represents missing values. The colour distances between the classes, computed according to the $\Delta$E 2000 metric by the International Commission on Illumination (CIE)\footnote{The deltas were computed using Color.js (\url{https://colorjs.io/}, last accessed: May 21, 2021).} were: A-B: 8.95, B-C: 7.68, C-D: 10.5. That is an average of 9. A value of 9 for the $\Delta$E$_{00}$ is well beyond the values for `Just Noticeable Difference' mentioned in the literature (0.5 in \cite{Yang2012} and 0.6 in \cite{Linhares2008}). It remains also far from 2, the (not recommended) minimum mentioned in \citep{Brychtova2015}. \textcolor{black}{Thus, the experiment was still relatively far from the lower bound of choropleth maps' effectiveness.}

\begin{table}[h!]
\caption{colours used during the study. All L A B values are based on an illuminant and reference angle for input / output values of D65 10°.}
\label{tab:colorused}
\resizebox{\textwidth}{!}{%
\begin{tabular}{|l|l|l|l|l|l|l|l|l|}
\hline
Class & colour                                                         & l     & a     & b     & R   & G   & B   & HEX    \\ \hline
Null & \fcolorbox{black}{class_null}{\rule{0pt}{6pt}\rule{6pt}{0pt}}\quad & 48.44 & 0.00 & 0.00 & 115 & 115 & 115 & 737373 \\ \hline
A     & \fcolorbox{black}{class_a}{\rule{0pt}{6pt}\rule{6pt}{0pt}}\quad & 99.40 & -4.28 & 12.44 & 255 & 255 & 229 & ffffe5 \\ \hline
B     & \fcolorbox{black}{class_b}{\rule{0pt}{6pt}\rule{6pt}{0pt}}\quad & 96.59 & -6.12 & 29.41 & 255 & 247 & 188 & fff7bc \\ \hline
C     & \fcolorbox{black}{class_c}{\rule{0pt}{6pt}\rule{6pt}{0pt}}\quad & 90.82 & -1.35 & 43.36 & 254 & 227 & 145 & fee391 \\ \hline
D     & \fcolorbox{black}{class_d}{\rule{0pt}{6pt}\rule{6pt}{0pt}}\quad & 82.46 & 9.61  & 64.34 & 254 & 196 & 79  & fec44f \\ \hline
\end{tabular}%
}
\end{table}

\subsection{Hypotheses}
There is limited previous work to build upon to formulate possible outcomes of the experiment given the novelty of choriented maps (colour+orientation on univariate attributes) and their usage for the first time on the experimental medium (mobile devices). Nonetheless, a few reasonable expectations can be formulated according to the best of the authors' knowledge at the moment of the experiment. For each hypothesis, the reason for the expected direction of the relationship is briefly mentioned.

\begin{itemize}

\item H1: choriented maps will perform better than choropleth. Two visual variables (colour+orientation) are better than one (colour), all other things being equal. Existing work on the combination of visual variables \citep{Nelson1999,Nelson2000} did not consider the interaction between colour and orientation. In the absence of preliminary insight, this is a reasonable assumption for univariate maps, which are the object of study in this work.

\item (Note: No theoretically-grounded statement can be made on the comparison between choriented maps and graduated symbol maps (GSM). Here, the argument that two are better than one cannot be applied. The comparison between choriented and graduated symbol maps is that of colour+orientation vs size, in contrast to the comparison colour+orientation vs colour for the encoding of geographic information.) 

\item H2: choriented mobile will perform better than choriented. There are several limitations of the visual processing system discussed in \citep{Franconeri2013}. One of these, called the crowding effect, refers to the situation where the presence of too many objects limits performance regarding the identification of specific objects. The assumption here is that filling in geometries with orientation lines all over the small screen (Figure \ref{fig:app_chorientedmap}) would lead to crowding effects that will decrease performance. Choriented mobile visualizations (Figure \ref{fig:app_chorientedmobile}) are much simpler in that respect and are not expected to suffer from any crowding effect.

\item H3: graduated symbol maps (GSM) will perform better than choropleth. This follows from \textcite{Garlandini2009}'s finding that size is better than colour value for change detection. Comparison can indeed be modelled as the problem of detecting similarity/differences between two entities and is thus a change detection problem. 

\item H4: the performance of the visualizations will be task-dependent (i.e. question-dependent). This has been often reported in previous studies, e.g. the comparison of interactive tables and interactive geovisualizations for question-answering on open data \citep{degbelo2018comparison}, the comparison of geovisualizations for the identification of correlations over space and time \citep{pena-araya2019}, and the comparison of visualizations to answer questions regarding route planning in a transportation scenario \citep{Kubicek2017}. This recurrent observation has not been formally labelled in the Cartography literature, but it may suffice here to state that this is a corollary of the `no free lunch theorem': ``if an algorithm performs well on a certain class of problems then it necessarily pays for that with degraded performance on the set of all remaining problems" \citep{Wolpert1997}\footnote{The algorithm in the quote above refers to computer-implementable instructions, but the idea becomes relevant to visual search through the metaphor `visual search is an algorithm' (algorithm for visual search = visualization type + user's visual search strategies).}.

\end{itemize}

`Perform better' here means surpass on at least one dimension  (efficiency, effectiveness, and confidence) while remaining at least as good on the others.



\subsection{Participants}
16 participants, six female and ten male, participated in the study. The ages ranged from 21 to 67 years, for an average of 31 years old (sd: 12). All participants stated that they own a mobile device with either Android (13 out of 16) or iOS (three out of 16) as the operating system. Participants were frequent users of maps in their mobile phones: (5/16) reported using applications that feature maps two to four times per week, (4/16) use them four to eight times per week, (6/16) use them more than eight times per week, and only (1/16) stated using mobile applications featuring maps once per week. Most participants (15/16) stated that they feel familiar or very familiar with choropleth maps. The remaining participant (1/16) was unfamiliar with choropleth maps. As to graduated symbol maps, (12/16) reported being familiar or very familiar with this visualization type; (3/16) stated that they are neither familiar nor unfamiliar, and (1/16) participant was not familiar with this visualization type.

\section{Preliminary results}
The analysis of the data was done using bootstrap confidence intervals (instead of p values alone) in line with recent calls in HCI research \citep{Dragicevic2016}. Confidence intervals estimate provide richer information than p-values alone. A confidence interval that does not include zero indicates statistical significance; the tighter the interval, and the farther from zero, the stronger the evidence. The R package bootES by \textcite{kirby2013boot} was used for the analysis. The next subsections present the results of the pairwise comparison of the different visualizations. The following applies to all tables. The first two columns of the result tables represent the two visualizations that are compared. A statistically significant difference between the two groups exists if the CI Low and CI High values, which describe the lower and upper bounds of the bootstrap confidence interval respectively, do not enclose zero. Positive values for the lower and upper confidence interval bounds mean that the visualization in the first column resulted in significantly higher values than the one in the second column (the same principle applies to negative confidence interval bounds). In some cases, the data points collected about the two visualizations are nearly identical. These did allow for bootstrapping since there are no differences and the result tables indicate this with grey cells. Significant differences between the two groups are highlighted in the tables with red borders and a coloured background. The bias is the difference between the mean of the resamples and the mean of the original sample. The SE (standard error) is the standard deviation of the resampled means. The number of resamples used in the analysis was N = 2000.


\subsection{Efficiency}
Table \ref{tab:efficiency} presents the efficiency results. The key takeaways are:  

\begin{itemize}

\item Choropleth vs Graduated Symbol: each visualization had a slight advantage for some tasks, but the advantages were not statistically significant.

\item Choropleth vs Choriented: here also, each visualization offered a minimal gain in task completion time, but the gains were not statistically significant. 

\item Choropleth vs Choriented Mobile: The choriented mobile visualization was faster for the identification of clusters (attributes-in-space). Though choropleth maps had a slight advantage for the remaining three tasks (frequency$|$attributes-in-space, trend$|$space-in-time and distribution$|$space-in-time), the advantages were not statistically significant. 

\item Choriented vs Graduated Symbol: each visualization offered a minimal gain in task completion time, but the gains were not statistically significant. Thus the two visualizations are comparable as far as time-on-task is concerned.

\item Choriented vs Choriented Mobile: here also, the choriented mobile visualization was faster for the identification of clusters (attributes-in-space). Though the differences for the remaining three tasks were not statistically significant, the tendency here was that choropleth maps were slightly faster for  space-in-time tasks (Trend/Distribution), and choriented faster for attributes-in-space tasks (Clusters/Frequency). 

\item Graduated Symbol vs Choriented Mobile: graduated symbol maps were faster for distribution tasks (space-in-time), but slower for cluster tasks (attributes-in-space). For trend and frequency tasks, choriented mobile maps had also a slight advantage, but the advantages were not statistically significant. 
\end{itemize}

\begin{table}[H]
\centering
\caption{
Bootstrapping result: Influence of the visualization types on the time needed when solving specific task types.
}
\arrayrulecolor{black}

\resizebox{\textwidth}{!}{%
\begin{tabular}{|l|l|l|l|l|l|l|l|} 
\hline
Vis. Type I                           & Vis. Type II                             & Task                                                                                     & Diff (seconds)                                                                                & CI Low                                                                               & CI High                                                                             & Bias                                                                               & SE                                                                                  \\ 
\hline
\multirow{4}{*}{Choropleth Map} & \multirow{4}{*}{GSM}               & Trend                                                                                    & -36.750                                                                             & -101.750                                                                             & 30.750                                                                              & -0.940                                                                             & 34.041                                                                              \\ 
\cline{3-8}
                                &                                    & Cluster                                                                                  & 1.500                                                                               & -29.000                                                                              & 57.331                                                                              & -0.441                                                                             & 21.515                                                                              \\ 
\cline{3-8}
                                &                                    & Distribution                                                                             & 20.750                                                                              & -4.000                                                                               & 52.750                                                                              & -0.099                                                                             & 13.802                                                                              \\ 
\cline{3-8}
                                &                                    & Frequency                                                                                & -27.250                                                                             & -62.750                                                                              & 7.500                                                                               & -0.237                                                                             & 18.204                                                                              \\ 
\hline
\multirow{4}{*}{Choropleth Map} & \multirow{4}{*}{Choriented Map}    & Trend                                                                                    & 6.500                                                                               & -27.750                                                                              & 39.750                                                                              & -0.149                                                                             & 17.445                                                                              \\ 
\cline{3-8}
                                &                                    & Cluster                                                                                  & 1.250                                                                               & -48.250                                                                              & 63.087                                                                              & 0.184                                                                              & 28.137                                                                              \\ 
\cline{3-8}
                                &                                    & Distribution                                                                             & -15.250                                                                             & -81.750                                                                              & 21.392                                                                              & 0.176                                                                              & 24.355                                                                              \\ 
\cline{3-8}
                                &                                    & Frequency                                                                                & -18.000                                                                             & -51.000                                                                              & 15.250                                                                              & 0.979                                                                              & 17.746                                                                              \\ 
\hline
\multirow{4}{*}{Choropleth Map} & \multirow{4}{*}{Choriented Mobile} & Trend                                                                                    & -1.000                                                                              & -30.000                                                                              & 31.633                                                                              & -0.311                                                                             & 15.965                                                                              \\ 
\hhline{|~~>{\arrayrulecolor{red}}------|}
                                &                                    & \multicolumn{1}{l!{\color{red}\vrule}}{{\cellcolor[rgb]{0.976,0.992,0.867}}Cluster}      & \multicolumn{1}{l!{\color{red}\vrule}}{{\cellcolor[rgb]{0.976,0.992,0.867}}37.250}  & \multicolumn{1}{l!{\color{red}\vrule}}{{\cellcolor[rgb]{0.976,0.992,0.867}}6.750}    & \multicolumn{1}{l!{\color{red}\vrule}}{{\cellcolor[rgb]{0.976,0.992,0.867}}90.750}  & \multicolumn{1}{l!{\color{red}\vrule}}{{\cellcolor[rgb]{0.976,0.992,0.867}}0.222}  & \multicolumn{1}{l!{\color{red}\vrule}}{{\cellcolor[rgb]{0.976,0.992,0.867}}21.959}  \\ 
\cline{3-8}
                                &                                    & Distribution                                                                             & -38.750                                                                             & -121.500                                                                             & 11.500                                                                              & 0.207                                                                              & 33.486                                                                              \\ 
\arrayrulecolor{black}\cline{3-8}
                                &                                    & Frequency                                                                                & -8.000                                                                              & -40.500                                                                              & 21.500                                                                              & 0.322                                                                              & 15.636                                                                              \\ 
\hline
\multirow{4}{*}{Choriented Map} & \multirow{4}{*}{GSM}               & Trend                                                                                    & -43.250                                                                             & -114.000                                                                             & 20.000                                                                              & 0.556                                                                              & 33.776                                                                              \\ 
\cline{3-8}
                                &                                    & Cluster                                                                                  & 0.250                                                                               & -34.750                                                                              & 35.000                                                                              & 0.218                                                                              & 18.084                                                                              \\ 
\cline{3-8}
                                &                                    & Distribution                                                                             & 36.000                                                                              & -0.750                                                                               & 102.165                                                                             & -0.494                                                                             & 24.710                                                                              \\ 
\cline{3-8}
                                &                                    & Frequency                                                                                & -9.250                                                                              & -52.250                                                                              & 23.994                                                                              & 0.171                                                                              & 19.362                                                                              \\ 
\hline
\multirow{4}{*}{Choriented Map} & \multirow{4}{*}{Choriented Mobile} & Trend                                                                                    & -7.500                                                                              & -35.750                                                                              & 28.000                                                                              & -0.125                                                                             & 15.953                                                                              \\ 
\hhline{|~~>{\arrayrulecolor{red}}------|}
                                &                                    & \multicolumn{1}{l!{\color{red}\vrule}}{{\cellcolor[rgb]{0.976,0.992,0.867}}Cluster}      & \multicolumn{1}{l!{\color{red}\vrule}}{{\cellcolor[rgb]{0.976,0.992,0.867}}36.000}  & \multicolumn{1}{l!{\color{red}\vrule}}{{\cellcolor[rgb]{0.976,0.992,0.867}}3.750}    & \multicolumn{1}{l!{\color{red}\vrule}}{{\cellcolor[rgb]{0.976,0.992,0.867}}73.384}  & \multicolumn{1}{l!{\color{red}\vrule}}{{\cellcolor[rgb]{0.976,0.992,0.867}}1.056}  & \multicolumn{1}{l!{\color{red}\vrule}}{{\cellcolor[rgb]{0.976,0.992,0.867}}18.725}  \\ 
\cline{3-8}
                                &                                    & Distribution                                                                             & -23.500                                                                             & -103.750                                                                             & 49.594                                                                              & -0.051                                                                             & 39.060                                                                              \\ 
\arrayrulecolor{black}\cline{3-8}
                                &                                    & Frequency                                                                                & 10.000                                                                              & -27.278                                                                              & 37.537                                                                              & -0.154                                                                             & 16.899                                                                              \\ 
\hline
\multirow{4}{*}{GSM}            & \multirow{4}{*}{Choriented Mobile} & Trend                                                                                    & 35.750                                                                              & -30.750                                                                              & 97.500                                                                              & -1.126                                                                             & 33.061                                                                              \\ 
\hhline{|~~>{\arrayrulecolor{red}}------|}
                                &                                    & \multicolumn{1}{l!{\color{red}\vrule}}{{\cellcolor[rgb]{0.976,0.992,0.867}}Cluster}      & \multicolumn{1}{l!{\color{red}\vrule}}{{\cellcolor[rgb]{0.976,0.992,0.867}}35.750}  & \multicolumn{1}{l!{\color{red}\vrule}}{{\cellcolor[rgb]{0.976,0.992,0.867}}25.750}   & \multicolumn{1}{l!{\color{red}\vrule}}{{\cellcolor[rgb]{0.976,0.992,0.867}}46.000}  & \multicolumn{1}{l!{\color{red}\vrule}}{{\cellcolor[rgb]{0.976,0.992,0.867}}-0.088} & \multicolumn{1}{l!{\color{red}\vrule}}{{\cellcolor[rgb]{0.976,0.992,0.867}}5.127}   \\ 
\hhline{>{\arrayrulecolor{black}}|~~>{\arrayrulecolor{red}}------|}
                                &                                    & \multicolumn{1}{l!{\color{red}\vrule}}{{\cellcolor[rgb]{0.976,0.992,0.867}}Distribution} & \multicolumn{1}{l!{\color{red}\vrule}}{{\cellcolor[rgb]{0.976,0.992,0.867}}-59.500} & \multicolumn{1}{l!{\color{red}\vrule}}{{\cellcolor[rgb]{0.976,0.992,0.867}}-151.599} & \multicolumn{1}{l!{\color{red}\vrule}}{{\cellcolor[rgb]{0.976,0.992,0.867}}-12.500} & \multicolumn{1}{l!{\color{red}\vrule}}{{\cellcolor[rgb]{0.976,0.992,0.867}}-0.435} & \multicolumn{1}{l!{\color{red}\vrule}}{{\cellcolor[rgb]{0.976,0.992,0.867}}33.217}  \\ 
\cline{3-8}
                                &                                    & Frequency                                                                                & 19.250                                                                              & -15.250                                                                              & 49.701                                                                              & 0.372                                                                              & 17.145                                                                              \\
\arrayrulecolor{black}\hline
\end{tabular}%
}
\label{tab:efficiency}
\end{table}

\subsection{Effectiveness}
Table \ref{tab:effectiveness} shows the effectiveness results. A higher score means that more questions were answered correctly. \textcolor{black}{The effectiveness score was computed by averaging over the answers given for the two questions belonging to a category (see Table \ref{tab:informationtasks}). Thus, if a participant answered both questions of a category correctly, they got an effectiveness score of 1, if they answered one out of two questions correctly 0.5 and if they failed to answer both questions correctly, they got a score of 0.} The key takeaway is that the different visualizations are largely comparable for the tasks considered.  

\begin{itemize}

\item Choropleth vs Graduated Symbol: Graduated Symbol Maps enabled the participants to produce slightly more accurate results for the identification of trends (space-in-time). The effectiveness of the participants was comparable for the remaining conditions, with a slight but non significant advantage for choropleth maps during the answering of cluster and distribution questions.

\item Choropleth vs Choriented: the effectiveness of the two visualizations were comparable. 

\item Choropleth vs Choriented Mobile: Participants were more effective in answering trend (space-in-time) questions using the choriented mobile visualization than the choropleth one. The effectiveness of the participants was comparable for the remaining three conditions. 

\item Choriented vs Graduated Symbol: The effectiveness of the participants was comparable for the two maps. Each had slight advantages for some tasks but the advantages were not statistically significant.

\item Choriented vs Choriented Mobile: the two are comparable as far as effectiveness is concerned. Choriented mobile had a slight advantage for trend (space-in-time) and frequency (attributes-in-space) tasks, but the advantages were not statistically significant.

\item Graduated Symbol vs Choriented Mobile: the two are comparable as far as effectiveness is concerned. Choriented mobile had a slight advantage for cluster (attributes-in-space) and distribution (space-in-time) tasks, but the advantages were not statistically significant.
\end{itemize}

\begin{table}[H]
\centering
\caption{
Bootstrapping result: Influence of the visualization types on the effectiveness when solving specific task types.
}
\arrayrulecolor{black}
\resizebox{\textwidth}{!}{%
 \begin{tabular}{|l|l|l|l|l|l|l|l|} 
\hline
Vis. Type I                     & Vis. Type II                       & Task                                                                          & Diff                                                                            & CI Low                                                                       & CI High                                                                      & Bias                                                                           & SE                                                                             \\ 
\hhline{|-->{\arrayrulecolor{red}}------|}
\multirow{4}{*}{Choropleth Map} & \multirow{4}{*}{GSM}               & \multicolumn{1}{l!{\color{red}\vrule}}{{\cellcolor[rgb]{0.992,1,0.898}}Trend} & \multicolumn{1}{l!{\color{red}\vrule}}{{\cellcolor[rgb]{0.992,1,0.898}}-0.125} & \multicolumn{1}{l!{\color{red}\vrule}}{{\cellcolor[rgb]{0.992,1,0.898}}-0.500} & \multicolumn{1}{l!{\color{red}\vrule}}{{\cellcolor[rgb]{0.992,1,0.898}}-0.125} & \multicolumn{1}{l!{\color{red}\vrule}}{{\cellcolor[rgb]{0.992,1,0.898}}0.004}  & \multicolumn{1}{l!{\color{red}\vrule}}{{\cellcolor[rgb]{0.992,1,0.898}}0.105}  \\ 
\cline{3-8}
                                &                                    & Cluster                                                                       & 0.125                                                                          & 0.000                                                                          & 0.250                                                                          & -0.004                                                                         & 0.107                                                                          \\ 
\arrayrulecolor{black}\cline{3-8}
                                &                                    & Distribution                                                                  & 0.125                                                                          & 0.000                                                                          & 0.250                                                                          & -0.000                                                                         & 0.106                                                                          \\ 
\hhline{|~~------|}
                                &                                    & Frequency                                                                     & \multicolumn{5}{l|}{{\cellcolor[rgb]{0.78,0.78,0.78}}}                                                                                                                                                                                                                                                                                                                                                             \\ 
\hline
\multirow{4}{*}{Choropleth Map} & \multirow{4}{*}{Choriented Map}    & Trend                                                                         & 0.000                                                                          & -0.375                                                                         & 0.125                                                                          & -0.003                                                                         & 0.153                                                                          \\ 
\hhline{|~~------|}
                                &                                    & Cluster                                                                       & \multicolumn{5}{l|}{{\cellcolor[rgb]{0.78,0.78,0.78}}}                                                                                                                                                                                                                                                                                                                                                             \\ 
\hhline{|~~------|}
                                &                                    & Distribution                                                                  & \multicolumn{5}{l|}{{\cellcolor[rgb]{0.78,0.78,0.78}}}                                                                                                                                                                                                                                                                                                                                                             \\ 
\cline{3-8}
                                &                                    & Frequency                                                                     & 0.125                                                                          & 0.000                                                                          & 0.250                                                                          & -0.000                                                                         & 0.108                                                                          \\ 
\hhline{|-->{\arrayrulecolor{red}}------|}
\multirow{4}{*}{Choropleth Map} & \multirow{4}{*}{Choriented Mobile} & \multicolumn{1}{l!{\color{red}\vrule}}{{\cellcolor[rgb]{0.992,1,0.898}}Trend} & \multicolumn{1}{l!{\color{red}\vrule}}{{\cellcolor[rgb]{0.992,1,0.898}}-0.125} & \multicolumn{1}{l!{\color{red}\vrule}}{{\cellcolor[rgb]{0.992,1,0.898}}-0.500} & \multicolumn{1}{l!{\color{red}\vrule}}{{\cellcolor[rgb]{0.992,1,0.898}}-0.125} & \multicolumn{1}{l!{\color{red}\vrule}}{{\cellcolor[rgb]{0.992,1,0.898}}-0.001} & \multicolumn{1}{l!{\color{red}\vrule}}{{\cellcolor[rgb]{0.992,1,0.898}}0.107}  \\ 
\hhline{>{\arrayrulecolor{black}}|~~>{\arrayrulecolor{red}}------>{\arrayrulecolor{black}}|}
                                &                                    & Cluster                                                                       & \multicolumn{5}{l|}{{\cellcolor[rgb]{0.78,0.78,0.78}}}                                                                                                                                                                                                                                                                                                                                                             \\ 
\hhline{|~~------|}
                                &                                    & Distribution                                                                  & \multicolumn{5}{l|}{{\cellcolor[rgb]{0.78,0.78,0.78}}}                                                                                                                                                                                                                                                                                                                                                             \\ 
\hhline{|~~------|}
                                &                                    & Frequency                                                                     & \multicolumn{5}{l|}{{\cellcolor[rgb]{0.78,0.78,0.78}}}                                                                                                                                                                                                                                                                                                                                                             \\ 
\hline
\multirow{4}{*}{Choriented Map} & \multirow{4}{*}{GSM}               & Trend                                                                         & -0.125                                                                         & -0.500                                                                         & 0.000                                                                          & -0.002                                                                         & 0.109                                                                          \\ 
\cline{3-8}
                                &                                    & Cluster                                                                       & 0.125                                                                          & ~0.000                                                                         & 0.250                                                                          & -0.001                                                                         & 0.105~                                                                         \\ 
\cline{3-8}
                                &                                    & Distribution                                                                  & 0.125                                                                          & 0.000                                                                          & 0.250                                                                          & -0.004                                                                         & 0.110                                                                          \\ 
\cline{3-8}
                                &                                    & Frequency                                                                     & -0.125                                                                         & -0.500                                                                         & 0.000                                                                          & -0.002                                                                         & 0.112                                                                          \\ 
\hline
\multirow{4}{*}{Choriented Map} & \multirow{4}{*}{Choriented Mobile} & Trend                                                                         & -0.125                                                                         & -0.500                                                                         & 0.000                                                                          & -0.001                                                                         & 0.110~                                                                         \\ 
\hhline{|~~------|}
                                &                                    & Cluster                                                                       & \multicolumn{5}{l|}{{\cellcolor[rgb]{0.78,0.78,0.78}}}                                                                                                                                                                                                                                                                                                                                                             \\ 
\hhline{|~~------|}
                                &                                    & Distribution                                                                  & \multicolumn{5}{l|}{{\cellcolor[rgb]{0.78,0.78,0.78}}}                                                                                                                                                                                                                                                                                                                                                             \\ 
\cline{3-8}
                                &                                    & Frequency                                                                     & -0.125                                                                         & -0.500                                                                         & 0.000                                                                          & -0.005                                                                         & 0.108                                                                          \\ 
\hline
\multirow{4}{*}{GSM}            & \multirow{4}{*}{Choriented Mobile} & Trend                                                                         & \multicolumn{5}{l|}{{\cellcolor[rgb]{0.78,0.78,0.78}}}                                                                                                                                                                                                                                                                                                                                                             \\ 
\cline{3-8}
                                &                                    & Cluster                                                                       & -0.125                                                                         & -0.500                                                                         & 0.000                                                                          & -0.002                                                                         & 0.106                                                                          \\ 
\cline{3-8}
                                &                                    & Distribution                                                                  & -0.125                                                                         & -0.500                                                                         & 0.000                                                                          & ~0.002                                                                         & 0.108                                                                          \\ 
\hhline{|~~------|}
                                &                                    & Frequency                                                                     & \multicolumn{5}{l|}{{\cellcolor[rgb]{0.78,0.78,0.78}}}                                                                                                                                                                                                                                                                                                                                                             \\
\hline
\end{tabular}%
}
\label{tab:effectiveness}
\end{table}

\subsection{Self-reported user confidence}
Table \ref{tab:confidence} shows the differences in confidence scores. \textcolor{black}{The confidence scores ranged for 1 to 5, with 5 indicating greater confidence}. A key takeaway here is that choropleth maps are not the most favourable option for maximized user confidence in answering trend (space-in-time) questions. They were consistently ranked lower than the other options available for that task. On the contrary, choriented mobile visualizations seem suitable for increased confidence in distribution (space-in-time) questions.

\begin{itemize}

\item Choropleth vs Graduated Symbol: users were more confident answering trend questions (space-in-time) with graduated symbol maps than with choropleth maps. On the contrary, they seemed more confident answering distribution questions (space-in-time) with choropleth maps than with graduated symbol maps. Choropleth maps had slight advantages for the remaining two tasks but the advantages were not statistically significant.

\item Choropleth vs Choriented: here also, users felt more confident answering trend questions (space-in-time) with choriented maps than with choropleth maps. On the contrary, they felt more confident answering distribution questions with choropleth maps than with choriented maps. 

\item Choropleth vs Choriented Mobile: users felt more confident answering trend (space-in-time) questions using the choriented mobile visualization than the choropleth visualization. The confidence of the participants was comparable for the remaining three tasks. 

\item Choriented vs Graduated Symbol: The reported confidence of the participants was comparable for the two maps. Each had slight advantages for some tasks but the advantages were not statistically significant.

\item Choriented vs Choriented Mobile: users were more confident in answering distribution (space-in-time) questions in the choriented mobile condition than in the choriented condition. The reported confidence was comparable in the remaining conditions. 

\item Graduated Symbol vs Choriented Mobile: here also, choriented mobile improved users' confidence when answering distribution (space-in-time) questions. The confidence scores were comparable in the remaining conditions.
\end{itemize}

\begin{table}[H]
\centering
\caption{
Bootstrapping result: Influence of the visualization types on the confidence when solving specific task types.
}
\label{tab:confidence}
\resizebox{\textwidth}{!}{%
\begin{tabular}{!{\color{black}\vrule}l!{\color{black}\vrule}l!{\color{black}\vrule}l!{\color{black}\vrule}l!{\color{black}\vrule}l!{\color{black}\vrule}l!{\color{black}\vrule}l!{\color{black}\vrule}l!{\color{black}\vrule}} 
\arrayrulecolor{black}\hline
Vis. Type I                     & Vis. Type II                       & Task                                                            & Diff                                                      & CI Low                                                    & CI High                                                   & Bias                                                                      & SE                                                        \\ 
\hhline{|-->{\arrayrulecolor{red}}------|}
\multirow{4}{*}{Choropleth Map} & \multirow{4}{*}{GSM}               & \multicolumn{1}{l|}{{\cellcolor[rgb]{0.996,1,0.8}}Trend}        & \multicolumn{1}{l|}{{\cellcolor[rgb]{0.996,1,0.8}}-1.250} & \multicolumn{1}{l|}{{\cellcolor[rgb]{0.996,1,0.8}}-1.875} & \multicolumn{1}{l|}{{\cellcolor[rgb]{0.996,1,0.8}}-0.875} & \multicolumn{1}{l|}{{\cellcolor[rgb]{0.996,1,0.8}}-0.002}                 & \multicolumn{1}{l|}{{\cellcolor[rgb]{0.996,1,0.8}}0.236}  \\ 
\cline{3-8}
                                &                                    & Cluster                                                         & 0.000                                                     & -0.750                                                    & 0.250                                                     & -0.005                                                                    & 0.305                                                     \\ 
\hhline{>{\arrayrulecolor{black}}|~~>{\arrayrulecolor{red}}------|}
                                &                                    & \multicolumn{1}{l|}{{\cellcolor[rgb]{0.996,1,0.8}}Distribution} & \multicolumn{1}{l|}{{\cellcolor[rgb]{0.996,1,0.8}}0.250}  & \multicolumn{1}{l|}{{\cellcolor[rgb]{0.996,1,0.8}}0.125}  & \multicolumn{1}{l|}{{\cellcolor[rgb]{0.996,1,0.8}}0.375}  & \multicolumn{1}{l|}{{\cellcolor[rgb]{0.996,1,0.8}}-0.001}                 & \multicolumn{1}{l|}{{\cellcolor[rgb]{0.996,1,0.8}}0.126}  \\ 
\cline{3-8}
                                &                                    & Frequency                                                       & 0.375                                                     & -0.125                                                    & 0.625                                                     & -0.000                                                                    & 0.205                                                     \\ 
\hhline{>{\arrayrulecolor{black}}|-->{\arrayrulecolor{red}}------|}
\multirow{4}{*}{Choropleth Map} & \multirow{4}{*}{Choriented Map}    & \multicolumn{1}{l|}{{\cellcolor[rgb]{0.996,1,0.8}}Trend}        & \multicolumn{1}{l|}{{\cellcolor[rgb]{0.996,1,0.8}}-0.875} & \multicolumn{1}{l|}{{\cellcolor[rgb]{0.996,1,0.8}}-1.625} & \multicolumn{1}{l|}{{\cellcolor[rgb]{0.996,1,0.8}}-0.125} & \multicolumn{1}{l|}{{\cellcolor[rgb]{0.996,1,0.8}}-0.006}                 & \multicolumn{1}{l|}{{\cellcolor[rgb]{0.996,1,0.8}}0.372}  \\ 
\cline{3-8}
                                &                                    & Cluster                                                         & -0.250                                                    & -1.000                                                    & 0.000                                                     & -0.003                                                                    & 0.215                                                     \\ 
\hhline{>{\arrayrulecolor{black}}|~~>{\arrayrulecolor{red}}------|}
                                &                                    & \multicolumn{1}{l|}{{\cellcolor[rgb]{0.996,1,0.8}}Distribution} & \multicolumn{1}{l|}{{\cellcolor[rgb]{0.996,1,0.8}}0.750}  & \multicolumn{1}{l|}{{\cellcolor[rgb]{0.996,1,0.8}}0.125}  & \multicolumn{1}{l|}{{\cellcolor[rgb]{0.996,1,0.8}}1.125}  & \multicolumn{1}{l|}{{\cellcolor[rgb]{0.996,1,0.8}}0.002} & \multicolumn{1}{l|}{{\cellcolor[rgb]{0.996,1,0.8}}0.282}  \\ 
\cline{3-8}
                                &                                    & Frequency                                                       & 0.625                                                     & -0.125                                   & 1.500                                                     & 0.000                                                                     & 0.425                                                     \\ 
\hhline{>{\arrayrulecolor{black}}|-->{\arrayrulecolor{red}}------|}
\multirow{4}{*}{Choropleth Map} & \multirow{4}{*}{Choriented Mobile} & \multicolumn{1}{l|}{{\cellcolor[rgb]{0.996,1,0.8}}Trend}        & \multicolumn{1}{l|}{{\cellcolor[rgb]{0.996,1,0.8}}-0.875} & \multicolumn{1}{l|}{{\cellcolor[rgb]{0.996,1,0.8}}-1.625} & \multicolumn{1}{l|}{{\cellcolor[rgb]{0.996,1,0.8}}-0.375} & \multicolumn{1}{l|}{{\cellcolor[rgb]{0.996,1,0.8}}0.000}                  & \multicolumn{1}{l|}{{\cellcolor[rgb]{0.996,1,0.8}}0.323}  \\ 
\cline{3-8}
                                &                                    & Cluster                                                         & 0.000                                                     & -0.750                                                    & 0.250                                                     & -0.007                                                                    & 0.249                                                     \\ 
\hhline{>{\arrayrulecolor{black}}|~~------|}
                                &                                    & Distribution                                                    & \multicolumn{5}{l!{\color{black}\vrule}}{{\cellcolor[rgb]{0.702,0.702,0.702}}}                                                                                                                                                                                                                                            \\ 
\cline{3-8}
                                &                                    & Frequency                                                       & 0.250                                                     & -0.250                                                    & \multicolumn{1}{l}{0.625}                                 & 0.003                                                                     & 0.233                                                     \\ 
\hline
\multirow{4}{*}{Choriented Map} & \multirow{4}{*}{GSM}               & Trend                                                           & -0.375                                                    & -1.375                                                    & 0.000                                                     & 0.004                                                                     & 0.324                                                     \\ 
\cline{3-8}
                                &                                    & Cluster                                                         & 0.250                                                     & 0.000                                                     & 0.500                                                     & 0.001                                                                     & 0.219                                                     \\ 
\cline{3-8}
                                &                                    & Distribution                                                    & -0.500                                                    & -1.250                                                    & 0.000                                                     & 0.002                                                                     & 0.314                                                     \\ 
\cline{3-8}
                                &                                    & Frequency                                                       & -0.250                                                    & -1.375                                                    & 0.375                                                     & 0.011                                                                     & 0.441                                                     \\ 
\hline
\multirow{4}{*}{Choriented Map} & \multirow{4}{*}{Choriented Mobile} & Trend                                                           & 0.000                                                     & -1.000                                                    & 0.625                                                     & -0.007                                                                    & 0.394                                                     \\ 
\cline{3-8}
                                &                                    & Cluster                                                         & 0.250                                                     & 0.000                                                     & 0.375                                                     & 0.001                                                                     & 0.127                                                     \\ 
\hhline{|~~>{\arrayrulecolor{red}}------|}
                                &                                    & \multicolumn{1}{l|}{{\cellcolor[rgb]{0.996,1,0.8}}Distribution} & \multicolumn{1}{l|}{{\cellcolor[rgb]{0.996,1,0.8}}-0.750} & \multicolumn{1}{l|}{{\cellcolor[rgb]{0.996,1,0.8}}-1.375} & \multicolumn{1}{l|}{{\cellcolor[rgb]{0.996,1,0.8}}-0.375} & \multicolumn{1}{l|}{{\cellcolor[rgb]{0.996,1,0.8}}0.014}                  & \multicolumn{1}{l|}{{\cellcolor[rgb]{0.996,1,0.8}}0.284}  \\ 
\cline{3-8}
                                &                                    & Frequency                                                       & -0.375                                                    & -1.500                                                    & 0.375                                                     & -0.005                                                                    & 0.463                                                     \\ 
\arrayrulecolor{black}\hline
\multirow{4}{*}{GSM}            & \multirow{4}{*}{Choriented Mobile} & Trend                                                           & 0.375                                                     & -0.250                                                    & 0.750                                                     & 0.004                                                                     & 0.270                                                     \\ 
\cline{3-8}
                                &                                    & Cluster                                                         & 0.000                                                     & -0.875                                                    & 0.250                                                     & -0.002                                                                    & 0.252                                                     \\ 
\hhline{|~~>{\arrayrulecolor{red}}------|}
                                &                                    & \multicolumn{1}{l|}{{\cellcolor[rgb]{0.996,1,0.8}}Distribution} & \multicolumn{1}{l|}{{\cellcolor[rgb]{0.996,1,0.8}}-0.250} & \multicolumn{1}{l|}{{\cellcolor[rgb]{0.996,1,0.8}}-0.500} & \multicolumn{1}{l|}{{\cellcolor[rgb]{0.996,1,0.8}}-0.125} & \multicolumn{1}{l|}{{\cellcolor[rgb]{0.996,1,0.8}}0.001}                  & \multicolumn{1}{l|}{{\cellcolor[rgb]{0.996,1,0.8}}0.127}  \\ 
\cline{3-8}
                                &                                    & Frequency                                                       & -0.125                                                    & -0.750                                                    & 0.375                                                     & -0.002                                                                    & 0.274                                                     \\
\arrayrulecolor{black}\hline
\end{tabular}%
}
\label{tab:confidence}
\end{table}

\subsection{Participants' subjective preference}
The question of participants' subjective preference touched upon all tasks. The two questions asked were `which visualization type was most suitable to solve the given tasks in your opinion'? and `which visualization type was least suitable to solve the given tasks'? \textcolor{black}{The users were also asked to provide the reasons for their choices.} Participants mostly preferred choropleth maps (11/16), followed by choriented maps (3/11), and graduated symbol maps/choriented mobile maps (1/11 each). Visualizations they liked less were graduated symbol maps (6/11), followed by choriented map/choriented mobile (5/11 each). \textcolor{black}{Key advantages mentioned were as follows: \textit{choropleth maps}: colour representation is easy to get an overview and it is the visualization that is ``easiest" to use; \textit{choriented maps}: better than choriented mobile because no overlap of markers happens. Disadvantages brought forth were: \textit{graduated symbol maps}: comparing circles is challenging and these circles overlap when the map is zoomed out; \textit{choriented maps}: two separate dimensions have to be attended to (or ``too much going on" as one participant put it); \textit{choriented mobile}: the small size of the symbols, the fact that markers overlap when the map is zoomed out, and the lack of highlighting of country borders.} In sum, the highlighting (or not) of the country borders, the simplicity of the visualization, the size of the symbols, and at last whether or not symbols overlap when zoomed out tipped participants' overall preference. These remarks explain why choropleth maps had some slight advantages as regards task completion time in some of the questions. Participants were thus faster, but not necessarily more accurate or even confident about their answers.

\subsection{Impact of participants' background}
The influence of participants' backgrounds on the dependent variables was also analyzed. The participants shared similar levels of knowledge when it came to the geographic locations of European countries and the usage and interpretation of the commonly used visualization types for SDG data. They also used mobile devices for map reading tasks similarly often. The few remaining variables for testing potential differences were thus: the gender, the operating system of the currently owned mobile phone (Android vs iOS), the daily mobile device usage, and the weekly usage of applications featuring maps. The tests were made for the four visualization types and the four task types considered during the experiment.

\begin{itemize}
    \item Gender: the differences regarding the four visualization types were mostly not significant (though a general tendency here was that male participants were slightly faster, more effective, and more confident across the use of all types of maps). An intriguing observation was that male participants were significantly faster and more effective using choriented maps than female participants. The performance on the task types (trends, cluster, frequency and distribution) was comparable. 
    \item Operating system: the results were inconclusive. 
    \item Daily device usage: this variable did not seem to have a significant impact on the results, except that participants using their device more often were slightly faster at using choriented mobile maps.
    \item Weekly usage of applications featuring maps: this variable did not have a significant impact on the results.
\end{itemize}

Overall, participants' background did not affect their interaction with the visualizations and their performance during the task completion. The gender effects observed are warranting a subsequent study for detailed investigation.

\section{Discussion}
Table \ref{tab:resultssummary} summarizes the results presented in the previous section. The four hypotheses are now revisited. As a reminder, `perform better' meant surpass on at least one dimension (efficiency, effectiveness, and confidence) while remaining at least as good on the others. To avoid data fishing, the interpretation of the results is conservative, that is, a direction is established at this point if it remained consistent across the three dimensions considered (efficiency, effectiveness, and confidence). Else, the visualizations are deemed to be comparable. 

\begin{itemize}

\item H1: choriented maps will perform better than choropleth. The observations in the study suggest that choriented maps perform better for the identification of space-in-time trends and the hypothesis is \textbf{supported} for this task. On the contrary, the statement does not hold for space-in-time distribution questions. Choropleth maps performed better here, with all three dimensions considered. The two visualizations appear to be largely comparable for the remaining two tasks considered: clusters$|$attributes-in-space and frequency$|$attributes-in-space. 

\item H2: choriented mobile will perform better than choriented. Despite some slight improvement in efficiency (cluster identification) and confidence (distributions identification), the improvements are not consistent across all three dimensions for the two tasks. Hence, this hypothesis is \textbf{not supported}. Until further evidence becomes available, choriented and choriented mobile visualizations can be considered comparable. The crowding effect seemed minimal in the current scenario.

\item H3: graduated symbol maps will perform better than choropleth. The hypothesis is \textbf{supported} for space-in-time trends identification. Graduated symbol maps resulted in users completing the trends identification tasks faster (even though not statistically significant), more effective (statistically significant), and more confidently (statistically significant). On the contrary, the data suggests that the statement does not hold for space-in-time distribution tasks. Indeed, though graduated symbol maps were faster for these tasks, choropleth maps led to improvements in confidence values and effective answers to the distribution tasks.

\item H4: the performance of the visualizations will be task-dependent. H4 is \textbf{supported}. H1 is supported for trends, not distributions; H3 is supported also for trends, and not distributions.
\end{itemize}

Hypotheses were formulated for three comparisons in Table \ref{tab:resultssummary}. The lessons learned for the remaining three comparisons are now summarized.

\begin{itemize}
\item Choriented mobile perform better than choropleth maps. In particular, choriented mobile maps result in gains in the efficiency and confidence of users during the identification of space-in-time trends. Choriented mobile maps also appear to perform better than choropleth maps for the identification of clusters (attributes-in-space), leading here to efficiency gains, without any loss on effectiveness or confidence.
\item Choriented mobile perform better than graduated symbol maps for the identification of clusters (attributes-in-space), leading here to efficiency gains, without any loss on effectiveness or confidence. As far as distribution (space-in-time) tasks are concerned, the two can be considered comparable until further evidence becomes available.
\item Choriented maps and graduated symbol maps are comparable across all four tasks considered.
\item All four visualization types are comparable for the answering of frequency (attribute-in-space) questions.
\end{itemize}

The questions were answered based on the average colour distance 9 (see Table \ref{tab:colorused}). Increasing this distance will likely improve the performance of choropleth maps; decreasing it will decrease the performance of choropleth maps. So, the conclusions touching on choropleth maps are only applicable for average colour distances of 9 and lower. Also, since only four classes were compared in this study, the scope of the conclusions is limited to four classes or lower.

\begin{table}[H]
\caption{Summary of the results. The approximation sign indicates that the two visualizations are comparable. A colour filling a whole cell indicates that the visualization performs better on that dimension (light blue: choropleth map, dark blue: graduated symbol map, light green: choriented, dark green: choriented mobile). Two colours in a cell indicate that the visualization was better for a task and worse on another task.}
\label{tab:resultssummary}
\resizebox{\textwidth}{!}{%
\begin{tabular}{|l|l|c|c|c|l|}
\hline
 &
   &
  \multicolumn{1}{l|}{Efficiency} &
  \multicolumn{1}{l|}{Effectiveness} &
  \multicolumn{1}{l|}{Confidence} &
  Notes \\ \hline
\cellcolor[HTML]{a6cee3} &
  \cellcolor[HTML]{1f78b4} &
   &
  \cellcolor[HTML]{1f78b4} &
  \cellcolor[HTML]{1f78b4} &
  Effe: trends; Conf (GSM): trends \\ \cline{5-6} 
\multirow{-2}{*}{\cellcolor[HTML]{a6cee3}Choropleth Map} &
  \multirow{-2}{*}{\cellcolor[HTML]{1f78b4}GSM} &
  \multirow{-2}{*}{$\approx$} &
  \multirow{-2}{*}{\cellcolor[HTML]{1f78b4}} &
  \cellcolor[HTML]{a6cee3} &
  Conf (Choro): distribution \\ \hline
\cellcolor[HTML]{a6cee3} &
  \cellcolor[HTML]{b2df8a} &
   &
   &
  \cellcolor[HTML]{b2df8a} &
  Conf (Chori): trends \\ \cline{5-6} 
\multirow{-2}{*}{\cellcolor[HTML]{a6cee3}Choropleth Map} &
  \multirow{-2}{*}{\cellcolor[HTML]{b2df8a}Choriented Map} &
  \multirow{-2}{*}{$\approx$} &
  \multirow{-2}{*}{$\approx$} &
  \cellcolor[HTML]{a6cee3} &
  Conf (Choro):  distribution \\ \hline
\cellcolor[HTML]{a6cee3} &
  \cellcolor[HTML]{33a02c} &
  \cellcolor[HTML]{33a02c} &
  \cellcolor[HTML]{33a02c} &
  \cellcolor[HTML]{33a02c} &
  Effi: clusters \\ \cline{6-6} 
\multirow{-2}{*}{\cellcolor[HTML]{a6cee3}Choropleth Map} &
  \multirow{-2}{*}{\cellcolor[HTML]{33a02c}Choriented Mobile} &
  \multirow{-2}{*}{\cellcolor[HTML]{33a02c}} &
  \multirow{-2}{*}{\cellcolor[HTML]{33a02c}} &
  \multirow{-2}{*}{\cellcolor[HTML]{33a02c}} &
  Effe: trends; Conf: trends \\ \hline
\cellcolor[HTML]{b2df8a} &
  \cellcolor[HTML]{1f78b4} &
   &
   &
   &
  - \\ \cline{6-6} 
\multirow{-2}{*}{\cellcolor[HTML]{b2df8a}Choriented Map} &
  \multirow{-2}{*}{\cellcolor[HTML]{1f78b4}GSM} &
  \multirow{-2}{*}{$\approx$} &
  \multirow{-2}{*}{$\approx$} &
  \multirow{-2}{*}{$\approx$} &
  - \\ \hline
\cellcolor[HTML]{b2df8a} &
  \cellcolor[HTML]{33a02c} &
  \cellcolor[HTML]{33a02c} &
   &
  \cellcolor[HTML]{33a02c} &
  Effi: clusters \\ \cline{6-6} 
\multirow{-2}{*}{\cellcolor[HTML]{b2df8a}Choriented Map} &
  \multirow{-2}{*}{\cellcolor[HTML]{33a02c}Choriented Mobile} &
  \multirow{-2}{*}{\cellcolor[HTML]{33a02c}} &
  \multirow{-2}{*}{$\approx$} &
  \multirow{-2}{*}{\cellcolor[HTML]{33a02c}} &
  Conf: distribution \\ \hline
\cellcolor[HTML]{1f78b4} &
  \cellcolor[HTML]{33a02c} &
  \cellcolor[HTML]{33a02c} &
   &
  \cellcolor[HTML]{33a02c} &
  Effi (ChoriM): clusters; Conf: distribution \\ \cline{3-3} \cline{6-6} 
\multirow{-2}{*}{\cellcolor[HTML]{1f78b4}GSM} &
  \multirow{-2}{*}{\cellcolor[HTML]{33a02c}Choriented Mobile} &
  \cellcolor[HTML]{1f78b4} &
  \multirow{-2}{*}{$\approx$} &
  \multirow{-2}{*}{\cellcolor[HTML]{33a02c}} &
  Effi (GSM): distribution \\ \hline
\end{tabular}%
}
\label{tab:resultssummary}
\end{table}

\subsection{Implications}

\subsubsection{Theoretical implications}
The impact of visual variables on graphical perception has been subject to investigations in previous work. \textcite{Garlandini2009} focused on the task of change detection in general and pointed out that size is more effective than colour value as pointed above. \textcite{Cleveland1984} focused on colour saturation and pointed out that it is less effective (i.e. accurate) than size for point comparison (i.e. finding the relative position of two data points). \textcite{Albers2014} investigated the use of colour (hue and value), as well as location as visual variables for point comparison (minima, maxima, range), and summary comparison (average, spread, outlier). They reported that colour supports summary comparison. In this study, the data taken as a starting point is classed numerical data. Thus, the comparison done is of type `point comparison' and more specifically \textit{range comparison}, i.e. finding out whether the attribute value of a given country falls within a certain range. Beyond the accuracy of answers (i.e. effectiveness) as the performance measure used, the work has also collected data on efficiency (i.e. how fast users produced an answer) and confidence. Triangulating confidence and effectiveness observations: 

\begin{itemize}
    \item Colour value and size are comparable as to selectivity of symbols for range comparison tasks on classed numerical data.
    \item Combining colour+orientation improves the selectivity of symbols during range comparison tasks on classed numerical data. (Compare the rows Choropleth vs GSM, and Choriented vs GSM in Table \ref{tab:resultssummary}).
    \item Combining colour+orientation offers similar selectivity performance as size on classed numerical data (row Choriented Map vs GSM in Table \ref{tab:resultssummary}).
    \item Combining colour+orientation does not systematically lead to greater selectivity in comparison with colour alone on classed numerical data. Visualization design and tasks may influence the selectivity gains (rows Choropleth vs Choriented and Choropleth vs Choriented Mobile in Table \ref{tab:resultssummary}).
\end{itemize}

Arguably, these statements remain preliminary at this point, but their key value is the provision of working hypotheses \citep{chamberlin1890method} for further empirical research on the graphical perception of map symbols.

\subsubsection{Practical implications}
As noted by \textcite{Kraak2020}, ``maps make it possible to see trends and make comparisons between different areas and over different time periods''. The current study has reminded that not all types of maps are equally effective at supporting these tasks. In particular, the following recommendations can be made. Again, the recommendations are preliminary and subject to the confirmation of the results through replication studies. They are also only applicable to mobile devices (medium used in this work) and classed numerical data (which were used throughout the study). 

\begin{itemize}
\item For trends (space-in-time) and cluster (attribute-in-space) tasks: choriented mobile maps should be used. They have consistently performed best than the other three visualizations investigated in the study.  
\item For trends (space-in-time) tasks: developers of choropleth maps should be aware of the fact that there are not as effective as alternatives. Users were consistently less confident giving answers using them which indicates that there are prone to incorrect interpretations. This is an important insight since finding space-in-time trends (i.e. how have values for a given country changed over a time period) is a basic task during the exploration/familiarization stage with datasets, and choropleth maps are often the only option available for this task in the context of SDG data (see Table \ref{tab:tools_table}). 
\item For distribution (space-in-time) tasks: choropleth maps are a reasonable option, as they perform at least as good or better than other maps on the dimensions of efficiency, effectiveness, and confidence. 
\item For frequency tasks: any of the four types of maps can be used. 
\end{itemize}

Empirically derived guidelines for the design of effective visualizations for geographic data have remained scarce and the list above can serve as a starting point for formulating these. Later, when they become consolidated, they can be translated into rules and/or models for the generation and optimization of intelligent geovisualizations \citep{degbelo2018intelligent}, for SDG data and beyond.


\subsection{Limitations and future work}
A limitation of the study is the limited number of participants and their relatively homogeneous background. The confidence in the results of this work can only increase as they are replicated in future studies. In addition, at least three other types of limitations can be mentioned: technical, study design, and application domain. \textcolor{black}{ From the technical point of view, the use of Mapbox GL as a web mapping library comes with the limitation that, at the moment of the writing, Mapbox GL still does not support non-mercator projections\footnote{See the full discussion about the issue at \url{https://github.com/mapbox/mapbox-gl-js/issues/3184} (last accessed: June 03, 2021).}. Though the conclusions of the work are not affected since all participants were exposed to the same stimulus across tasks during the study, being able to include equal-area projections that are more suitable for comparison purposes is desirable, and the switch to an equal-area projection can be easily done in the prototype once the feature becomes available.} From the study design point of view, strategy is an important dimension of problem-solving with maps but was not explicitly accounted for. It follows that the implicit assumption of the work was that all users would always use the `best' strategy per visualization for the information-gathering tasks. This is arguably simplistic as previous work (e.g. \cite{Winn1990}) discussed. \textcolor{black}{Repeating the study with more users and explicitly collecting information about their strategies during the study (e.g. via eye-tracking, or by systematically encoding their action sequences using the touch-level model for mobile devices \citep{rice2014touch}) would be necessary to confirm the efficiency and effectiveness observations made during the study.} Finally, from the application domain point of view, though the comparison of SDG values between countries is important, it must be stressed that it is not without its own limitations. Indeed, seeing that a country A has a higher value on a few SDG indicators than a country B might lead users to the overgeneralization that A is doing better than B when it comes to sustainable development. Given that the visualizations say nothing about the priorities of the countries in a year (or previous years), users should be educated to refrain from these hasty generalizations. Appropriate warnings and notifications could be included in the next version of the prototype to address that point. 

In addition to the ideas above that would shed further insight on the hypotheses investigated in this work, future work could also assess the perceptual properties of choriented visualizations represented as small multiples. While the results observed in the study apply to interaction scenarios, considering small multiples will be useful to assess the performance of the visualizations for animation scenarios. Another interesting area of investigation is the development of techniques to assess the visual distance between symbols for variables other than colour (e.g. size) and the visual distance between symbols with two variables (e.g. colour+orientation). These techniques are currently lacking and this has prevented a detailed documentation of all visual distance parameters of the study. \textcolor{black}{For instance, there is some evidence in the literature that the spatial distance between symbols impacts the performance of map reading tasks on graduated symbol maps \citep{Cybulski2020a}, and of colour discrimination tasks on choropleth maps \citep{Brychtova2017}. Measuring spatial distance between symbols on an interactive mobile map is a challenge in its own right and was thus not tackled during this work.} \textcolor{black}{A further area of investigation is the comparison of colour+orientation with other alternatives for redundant symbolization, for instance colour+size, colour+texture or colour+transparency (used e.g. for value-by-alpha maps, see \cite{Roth2010}). Next, identifying the higher bound of effectiveness of the colour+orientation approach (i.e. up to which number of classes are symbols resulting from the combination still distinguishable?) is also an important question. Finally, it would be interesting to replicate the study, particularly on a Desktop setting, to learn about the effect of small screen sizes on the conclusions of the study.}


\section{Conclusion}
As \textcite{tufte1990envisioning} remarked: ``At the heart of quantitative reasoning is a single question: \textit{Compared to what?}'' (emphasis added). This quote stresses the necessity of developing visualizations that effectively support comparison, and understanding their respective merits. A practical domain where the question ``compared to what?'' (or rather ``compared to whom'') is helpful to put numbers into perspective is that of the Sustainable Development Goals (SDGs). SDG data are becoming gradually available and present a good opportunity for research at the intersection of visualization and comparison. In addition, mobile devices are increasingly used for information-gathering tasks. Visualizations should thus be optimized for consumption on mobile phones. This notwithstanding, ``a need remains for best practices grounded in empirical evidence with regards to mobile-specific visualization design''\citep{Brehmer2020}.

To address this gap, this article has introduced Choriented Maps and Choriented Mobile, two new types of visualizations that use orientation and colour concurrently to encode geographic data. They are both meant to assist in cases when distinguishing between adjacent classes is challenging. Choriented maps fill geometries with orientation+colour patterns, and choriented mobile maps use small squares filled with orientation+colour patterns to convey geographical knowledge. The impact of the two visualizations on performance, accuracy, and users' confidence in giving answers was tested in a user study involving attribute-in-space and space-in-time comparison tasks on mobile devices. Choriented Mobile visualizations have consistently performed better than Choriented Map visualizations, Graduated Symbol Maps, and Choropleth maps for trends (space-in-time) and cluster (attribute-in-space) tasks and are thus a promising technique to support sensemaking of SDG data on mobile devices.

\section*{Supplementary material}
The applications mentioned in Section \ref{sec:relatedwork} were last accessed on May 13, 2021. The links are provided below. 

\begin{itemize}
    \item SDGs \& me:  \url{https://ec.europa.eu/eurostat/cache/digpub/sdgs/index.html}.
    \item World Development Indicators: \url{https://databank.worldbank.org/source/world-development-indicators}.
    \item SDGs Dashboard: \url{https://www.sdgsdashboard.org/}.
     \item SDG Index Dashboard: \url{https://www.sdgindex.org/}.
     \item Our World in Data: \url{https://ourworldindata.org/}.
    \item VizHub Health data: \url{https://vizhub.healthdata.org/sdg/}.
\end{itemize}

\section*{Acknowledgements}
We thank Christian Kray for his feedback during this work.

\section*{Conflict of interest}
The authors declare that they have no conflict of interest.

\section*{Biographical note}
Viktor Gorte holds a Master of Science degree in Geoinformatics from the Westfälische Wilhelms-Universität Münster. Since the beginning of his studies, he was interested in innovative ways to visualize and analyze spatial information. In his bachelor thesis, he compared drone and satellite images and converted them to 3D models. After finishing his bachelor's degree, he got more involved with web and mobile technologies and has worked on numerous applications that facilitate the usage and understanding of spatial information. Viktor currently works as a software developer where he uses state of the art technologies to make spatial information more accessible to everyone.\\

\noindent Auriol Degbelo is a Privatdozent at the University of Münster, Germany. His research interests include semantic integration of geospatial information, re-use of open government data, and interaction with geographic information. Past contributions of his work include a theory of spatial and temporal resolution of sensor observations, semantic APIs for the retrieval of open government data, and a semi-automatic approach for the creation of thematic web maps.


\printbibliography
\end{document}